\documentclass[12pt,preprint]{aastex}
\newcommand{\ergcms}{\ensuremath{\mathrm{erg}\,\mathrm{cm}^{-2}\,\mathrm{s}^{-1}}}%
\newcommand{\cms}{cm$^{-2}$\,s$^{-1}$}
\newcommand{\hess}{\textsc{H.E.S.S.}}
\newcommand{\fer}{{\sl {\it Fermi}}}
\newcommand{\fla}{\fer-LAT}
\newcommand{\fvar}{\mathrm{F}_{\mathrm{var}}}
\newcommand{\gr}{$\gamma$-ray}
\newcommand{\grs}{$\gamma$-rays}
\newcommand{\dgr}{\ensuremath{^\circ}}
\newcommand{\pg}{PG~1553+113}
\newcommand{\eqg}{E$_{\rm QG}$}
\newcommand{\bestz}{$z=0.49\pm0.04$}

\shorttitle{The 2012 flare of \pg\ seen with H.E.S.S. and \fla}
\shortauthors{The H.E.S.S. Collaboration}
\slugcomment{To be submitted to Astrophysical Journal}
\keywords{Galaxies: active -- BL Lacertae objects: Individual:
  \pg\ -- Gamma rays: observations -- Quantum Gravity -- Lorentz invariance breaking}
\usepackage{lineno}
\usepackage{graphicx}
\usepackage{natbib}
\usepackage{amsmath}
\begin{document}
\title{The 2012 flare of \pg\ seen with H.E.S.S. and \fla : Constraints on the source redshift 
and Lorentz invariance violation }  
\author{H.E.S.S. Collaboration,
A.~Abramowski \altaffilmark{1},
F.~Aharonian \altaffilmark{2,3,4},
F.~Ait Benkhali \altaffilmark{2},
A.G.~Akhperjanian \altaffilmark{5,4},
E.O.~Ang\"uner \altaffilmark{6},
M.~Backes \altaffilmark{7},
S.~Balenderan \altaffilmark{8},
A.~Balzer \altaffilmark{9},
A.~Barnacka \altaffilmark{10,11}
% ...
}

% ...
\altaffiltext{1}{Universit\"at Hamburg, Institut f\"ur Experimentalphysik, Luruper Chaussee 149, D 22761 Hamburg, Germany }
\altaffiltext{2}{Max-Planck-Institut f\"ur Kernphysik, P.O. Box 103980, D 69029 Heidelberg, Germany }
\altaffiltext{3}{Dublin Institute for Advanced Studies, 31 Fitzwilliam Place, Dublin 2, Ireland }
\altaffiltext{4}{National Academy of Sciences of the Republic of Armenia,  Marshall Baghramian Avenue, 24, 0019 Yerevan, Republic of Armenia  }
\altaffiltext{5}{Yerevan Physics Institute, 2 Alikhanian Brothers St., 375036 Yerevan, Armenia }
\altaffiltext{6}{Institut f\"ur Physik, Humboldt-Universit\"at zu Berlin, Newtonstr. 15, D 12489 Berlin, Germany }
\altaffiltext{7}{University of Namibia, Department of Physics, Private Bag 13301, Windhoek, Namibia }
\altaffiltext{8}{University of Durham, Department of Physics, South Road, Durham DH1 3LE, U.K. }
\altaffiltext{9}{GRAPPA, Anton Pannekoek Institute for Astronomy, University of Amsterdam,  Science Park 904, 1098 XH Amsterdam, The Netherlands }
\altaffiltext{10}{Obserwatorium Astronomiczne, Uniwersytet Jagiello{\'n}ski, ul. Orla 171, 30-244 Krak{\'o}w, Poland }
\altaffiltext{11}{now at Harvard-Smithsonian Center for Astrophysics,  60 Garden St, MS-20, Cambridge, MA 02138, USA }
% ...

\begin{abstract}
Very high energy (VHE, $E>$100~GeV) $\gamma$-ray flaring activity of the 
high-frequency peaked BL~Lac object \pg\ has been detected by the \hess\ 
telescopes. The flux of the source increased by a factor of 3 during the
nights of 2012 April 26 and 27 with respect to the archival measurements with
hint of intra-night variability. No counterpart of this event has been
detected in the \fla\ data. This pattern is consistent with VHE $\gamma$ ray
flaring being caused by the injection of ultrarelativistic particles, emitting
$\gamma$ rays at the highest energies. The dataset offers a unique opportunity
to constrain the redshift of this source at \bestz\ using a novel method based
on Bayesian statistics. The indication of intra-night variability is used to introduce a novel method to probe for a possible Lorentz Invariance Violation (LIV), and to set limits on the energy scale at which
Quantum Gravity (QG) effects causing LIV may arise. For the subluminal case,
the derived limits are $\textrm{E}_{\rm QG,1}>4.10\times 10^{17}$~GeV and 
$\textrm{E}_{\rm QG,2}>2.10\times 10^{10}$~GeV for linear and quadratic LIV
effects, respectively.
\end{abstract}

\section{Introduction}

Blazars are active galactic nuclei (AGN) with their jets closely 
aligned with the line of sight to the Earth \citep{THEO:Unification2}. 
Among their particularities is flux variability at all wavelengths on 
various time scales, from years down to (in some cases) minutes \citep{REF::MRK_FLARE_NATURE,REF::TEV_PKS_FLARE}. Flaring 
activity of blazars is of great interest for probing the source-intrinsic
physics of relativistic jets, relativistic particle acceleration and 
generation of high-energy radiation, as well as for conducting fundamental 
physics tests. On the one hand, exploring possible spectral variability 
between flaring and stationary states helps to understand the 
electromagnetic emission mechanisms at play in the jet. On the other 
hand, measuring the possible correlation between photon energies and 
arrival times allows one to test for possible Lorentz 
invariance violation (LIV) leading to photon-energy-dependent 
variations in the speed of light in vacuum.

Located in the Serpens Caput constellation, \pg\ was discovered by
\citet{Green:1986:PGCatalog}, who first classified it as a BL~Lac object. Later
the classification was refined to a high-frequency peaked BL~Lac object
\citep[HBL,][]{1995A&AS..109..267G}. \pg\ exhibits a high X-ray to radio
flux \citep[$\log(F_{\rm 2~keV}/F_{\rm 5~GHz})>-4.5$,]
[]{Osterman:2006:PG1553mwl}, which places it among the most extreme HBLs
\citep{Rector:2003:RadioHBL}. The object was observed in X-rays by multiple
instruments in different flux states. Its 2--10~keV energy flux ranges from $0.3
\times 10^{-11}\ \ergcms$ \citep {Osterman:2006:PG1553mwl} to $3.5 \times
10^{-11} \ergcms$ \citep {Reimer:2008:PG1553Suzaku} but no fast variability (in
the sub-hour time scale) has been detected so far. 

\pg\ was discovered at very high energies (VHE, $E>$100~GeV) by H.E.S.S.
\citep{2006A&A...448L..19A,2008A&A...477..481A} with a photon index of $\Gamma
=4.0\pm0.6$. At high energies (HE, 100~MeV$<E<$300~GeV) the source has been
detected by the \fla\ \citep {Fermi:2009:BrightSourceList,2010ApJ...708.1310A}
with a very hard photon index of $\Gamma = 1.68 \pm0.03$, making this object the
one with the largest HE -- VHE spectral break ($\Delta \Gamma\approx2.3$ ) ever
measured. No variability in \fla\ was found by \citet
{Fermi:2009:BrightSourceList,2010ApJ...708.1310A} on daily or weekly time scales, 
but using an extended data set of 17 months, \citet{Magic5years} reported
variability above 1~GeV with flux variations of a factor of $\sim 5$ on a yearly 
time scale.  

With 5 years of monitoring data of the MAGIC telescopes, \citet {Magic5years}
discovered variability in VHE $\gamma$ rays with only modest flux variations
(from 4 to 11~\% of the Crab Nebula flux). In addition to the high X-ray
variability, this behavior can be interpreted as evidence for Klein-Nishina
effects \citep {2010ApJ...708.1310A} in the framework of a synchrotron
self-Compton model. The source underwent VHE $\gamma$-ray flares in 2012 March
\citep{2012ATel.3977....1C} and April \citep{2012ATel.4069....1C}, detected by
the MAGIC telescopes. During the March flare, the source was at a flux level of
about 15\% of that of the Crab Nebula, while in April it reached $\approx 50\%$.
During those VHE $\gamma$-ray flares, also a brightening in X-ray, UV and optical 
wavelengths has been noticed by the MAGIC collaboration. A detailed study of the 
MAGIC telescopes and multi-wavelength data is in press \citep{2014arXiv1408.1975A}. The latter event triggered 
the \hess\ observations reported in this work.

Despite several attempts to measure it, the redshift of \pg\ still suffers from
uncertainties. Different attempts, including optical spectroscopy \citep
{2007A&A...473L..17T,2008A&A...477..481A} or comparisons of the HE and VHE
spectra of \pg\ \citep{2009arXiv0907.0157P,2013A&A...554A..75S}, were made.
Based on the assumption that the EBL-corrected VHE spectral index is equal 
to the \fla\ one, \citet{2009arXiv0907.0157P} derived an upper limit of $z<0.67$.
Comparing \pg\ statistically with other known VHE emitters and taking into 
account a possible intrinsic $\gamma$-ray spectral break through a simple
emission model, \citet{2013A&A...554A..75S} constrained the redshift to be 
below 0.64. The best estimate to-date was obtained by \citet {danforth_hubble/cos_2010} 
who found the redshift to be between 0.43 and 0.58 using far-ultraviolet spectroscopy.

This paper concentrates on the HE and VHE emission of \pg\ and is divided
as follow: Sections \ref{anahess} and \ref{anafermi} present the \hess\ and
\fla\ analyses. The discussion, in section \ref{discussion}, includes
the determination of the redshift using a novel method and the constraints
derived on LIV using a modified likelihood formulation. Throughout this
paper a $\Lambda$CDM cosmology with H$_0 = 70.4 \pm 1.4$
km\,s$^{-1}\,$Mpc$^{-1}$, $\Omega_m = 0.27 \pm 0.03$, $\Omega_\Lambda = 0.73 \pm
0.03$ from WMAP \citep {komatsu_seven-year_2011} is assumed.

\section{Data analysis}\label{ana}
\subsection{\hess\ observations and analysis}\label{anahess}

\hess\ is an array of five imaging atmospheric Cherenkov telescopes located in
the Khomas highland in Namibia ($23^\circ16'18''$~S, $16^\circ30'01''$~E), at
an altitude of 1800~m above sea level \citep{2004NewAR..48..331H}. The fifth
\hess\ telescope was added to the system in 2012 July and is not used in this 
work, reporting only on observations prior to that time.

\pg\ was observed with \hess\ in 2005 and 2006 \citep{2008A&A...477..481A}. No
variability was found in these observations, which will be referred to as the
``pre-flare'' data set in the following. New observations were carried out in
2012 April after flaring activity at VHE was reported by the MAGIC collaboration \citep[``flare'' data set,][]{2012ATel.4069....1C}.

The pre-flare data set is composed of $26.4$ live time hours of good-quality
data \citep{aha2006}. For the flare period, eight runs of $\sim 28$ minutes each
were taken during the nights of 2012 April 26 and 27, corresponding to $3.5$ hours
of live time. All the data were taken in wobble mode, for which the source is
observed with an offset of $0\fdg 5$ with respect to the center of the
instrument's field of view yielding an acceptance-corrected live time of
24.7 hours and 3.2 hours for the pre-flare and flare data sets, respectively.

Data were analyzed using the {\tt Model} analysis \citep{2009APh....32..231D}
with {\tt Loose cuts}. This method--based on the comparison of detected shower
images with a pre-calculated model--achieves a better rejection of hadronic
air showers and a better sensitivity at lower energies than analysis methods
based on Hillas parameters. The chosen cuts, best suited for sources with steep
spectra such as \pg\footnote{\pg\ has one of the steepest spectra measured at
VHE.}, require a minimum image charge of 40 photoelectrons,
which provides an energy threshold of $\rm \sim 217$~GeV for the pre-flare and
$\rm \sim 240$~GeV for the flare data set\footnote{The difference of energy threshold between the two data set is due to the changing observation conditions, e.g., zenith angle and optical efficiency.}. All the results presented in this
paper were cross-checked with the independent analysis chain described in
\citet{2011APh....34..858B}.

Events in a circular region (ON region) centered on the radio position
of the source, $\alpha_{\rm J2000} = 15^{\rm h}55^{\rm m}43.04^{\rm
s},\delta_{\rm J2000} =11^{\circ}11'24.4''
$ \citep{Green:1986:PGCatalog}, with a maximum squared angular
distance of $0.0125~{\rm deg}^2$, are used for the analysis. In order
to estimate the background in this region, the reflected
background method \citep{2007A&A...466.1219B} is used to define the
OFF regions. The excess of $\gamma$ rays in the ON region is statistically
highly significant \citep{Lima}: $21.5 \,\sigma$ for the pre-flare period and
$22.0 \, \sigma$ for the flare. Statistics are summarized
in Table~\ref{tab:hessNum}.

\begin{deluxetable}{lcccccccc}
\tablecaption{Summary of the statistics for both data sets (first column). The
second and third columns give the number of ON and OFF events. The 4th column
gives the ratio between ON and OFF exposures ($r$). The excess and the
corresponding significance are given, as well as the energy threshold and the
mean zenith angle of the source during the observations. The last column
presents the probability of the flux to be constant within the observations (see
text).\label{tab:hessNum}}

\tablewidth{0pt}
\tablehead{\colhead{Data set} & \colhead{ON} &\colhead{OFF} &\colhead{$r$} &\colhead{Excess}
&\colhead{Significance} &\colhead{$E_{\rm th}\,\mathrm{[GeV]}$}& Zenith angle &\colhead{$P^{\rm cst}_{\chi^2}$}}
\startdata
    Pre-Flare & 2205 & 13033 & 0.100 & 901.7 & 21.5 & 217 & $34^\circ$ & 0.77\\\hline
    Flare & 559 & 1593 & 0.105 & 391.2 & 22.0 & 240 & $52^\circ$ & $3.3 \times 10^{-3}$\\ 
\enddata
\end{deluxetable}

The differential energy spectrum of the VHE \gr\ emission has been derived using
a forward-folding method \citep{2001AA...374..895P}. For the observations prior
to 2012 April, a power law (PWL) model fitted to the data gives a $\chi^2$ of
$51.7$ for $40$ degrees of freedom (d.o.f., corresponding to a $\chi^2$
probability of $P_{\rm \chi^2}=0.10$). The values of the spectral parameters (see
Table~\ref{tab:specvhe}) are compatible with previous analyses by H.E.S.S.
covering the same period \citep{2008A&A...477..481A}. A log-parabola (LP)
model\footnote{The log-parabola is defined by $dN/dE = \Phi_0
\left(E/E_0\right)^{-a - b \log(E/E_0)}$.}, with a $\chi^2$ of $37.5$ for $39$
d.o.f. ($P_{\rm \chi^2}=0.54$), is found to be preferred over the PWL model at 
a level of $4.3 \, \sigma$ using the log-likelihood ratio test. Note that
systematic uncertainties, presented in Table~\ref{tab:specvhe}, have been
evaluated by \citet{aha2006} for the PWL model and using the jack-knife method
for the LP model. The jack-knife method consist in removing one run and redoing the analysis. This process is repeated for all runs.

For the flare data set, the log-parabola model does not significantly improve
the fit and the simple PWL model describes the data well, with a $\chi^2$
of $33.0$ for $23$ d.o.f. ($P_{\rm \chi^2}=0.08$). Table~\ref{tab:specvhe}
contains the integral fluxes above the reference energy of 300 GeV. The flux increased
by a factor of $\sim3$ in the flare data set compared to the pre-flare one with
no sign of spectral variations (when comparing power law fits for both data
sets).  The derived spectra and error contours for each data set are presented in
Fig.~\ref{fig:specvhe}, where the spectral points obtained from the cross-check
analysis are also plotted.

\newlength{\specwidth}
\setlength{\specwidth}{0.49\textwidth}
\newcommand{\includeSpec}[1]{\includegraphics[width=\specwidth]{#1}}

\begin{figure}[p]
\centering
\includeSpec{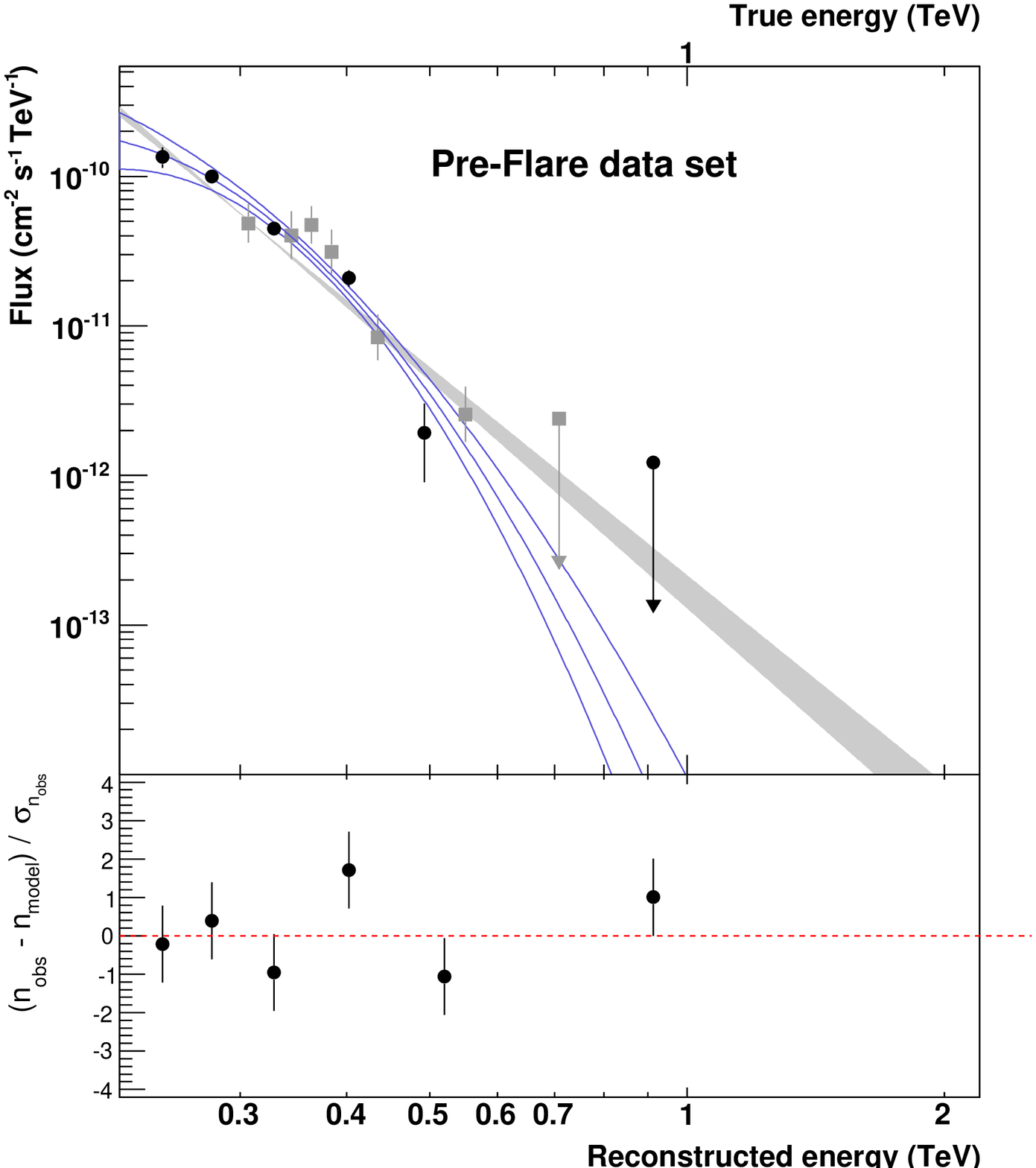}%
\includeSpec{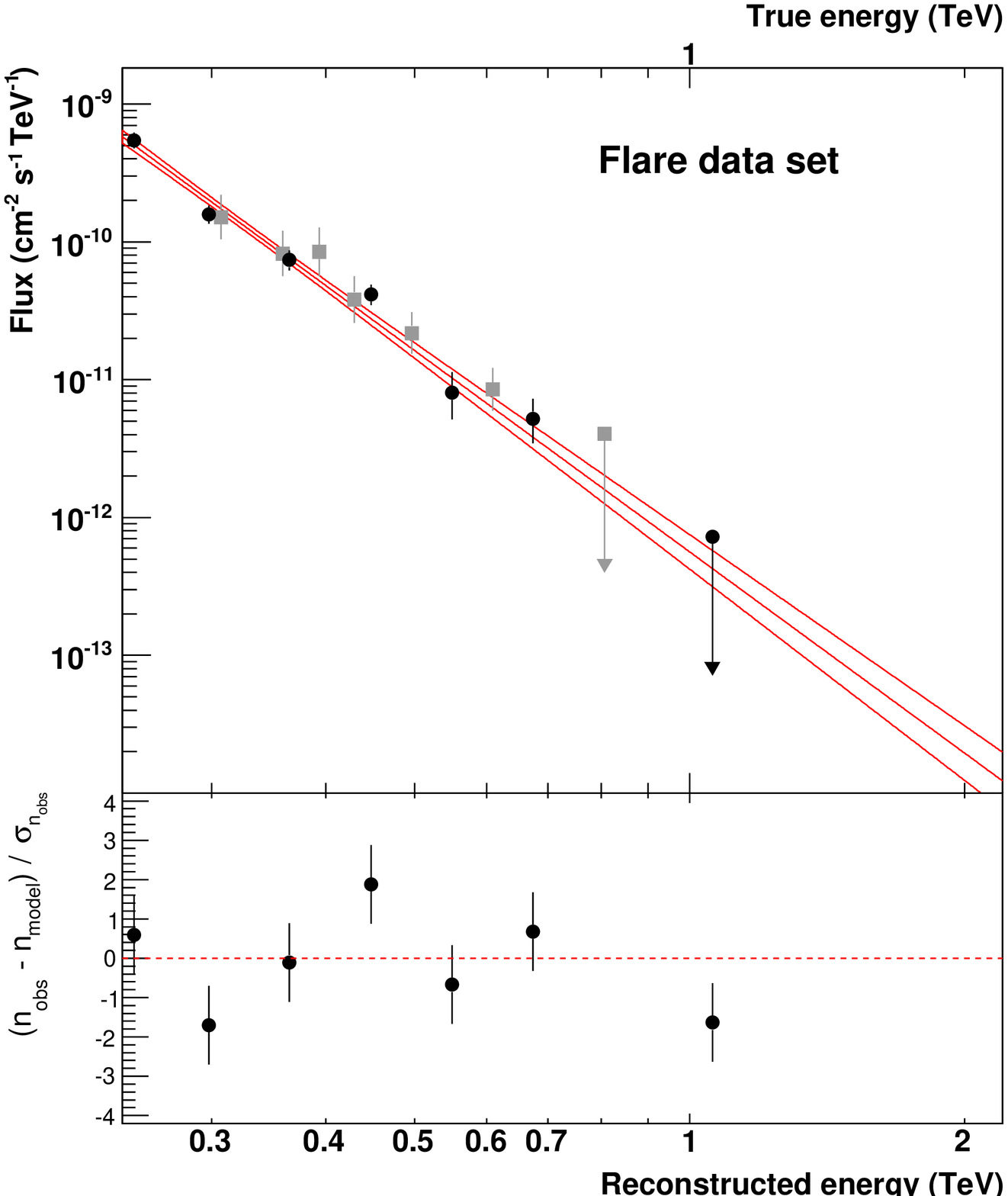}

\caption{Differential fluxes of \pg\ during the pre-flare (left) and
flare (right) periods. Error contours indicate the 68~\% uncertainty on the spectrum.
Uncertainties on the spectral points (in black) are given at $1 \, \sigma$ level, and upper
limits are computed at the 99~\% confidence level. The gray squares were
obtained by the cross-check analysis chain and are presented to visualize the match between both
analyses. The gray error contour on the left panel is the best-fit power law
model. The lower panels show the residuals of the fit, i.e. the difference between
the measured ($n_{\mathrm{obs}}$) and expected numbers of photons
($n_{\mathrm{model}}$), divided by the statistical error on the measured 
number of photons ($\sigma_{n_{\mathrm{obs}}}$).}
\label{fig:specvhe}
\end{figure}

\begin{figure}[ht]
		 \centering
     \includegraphics[width=\textwidth]{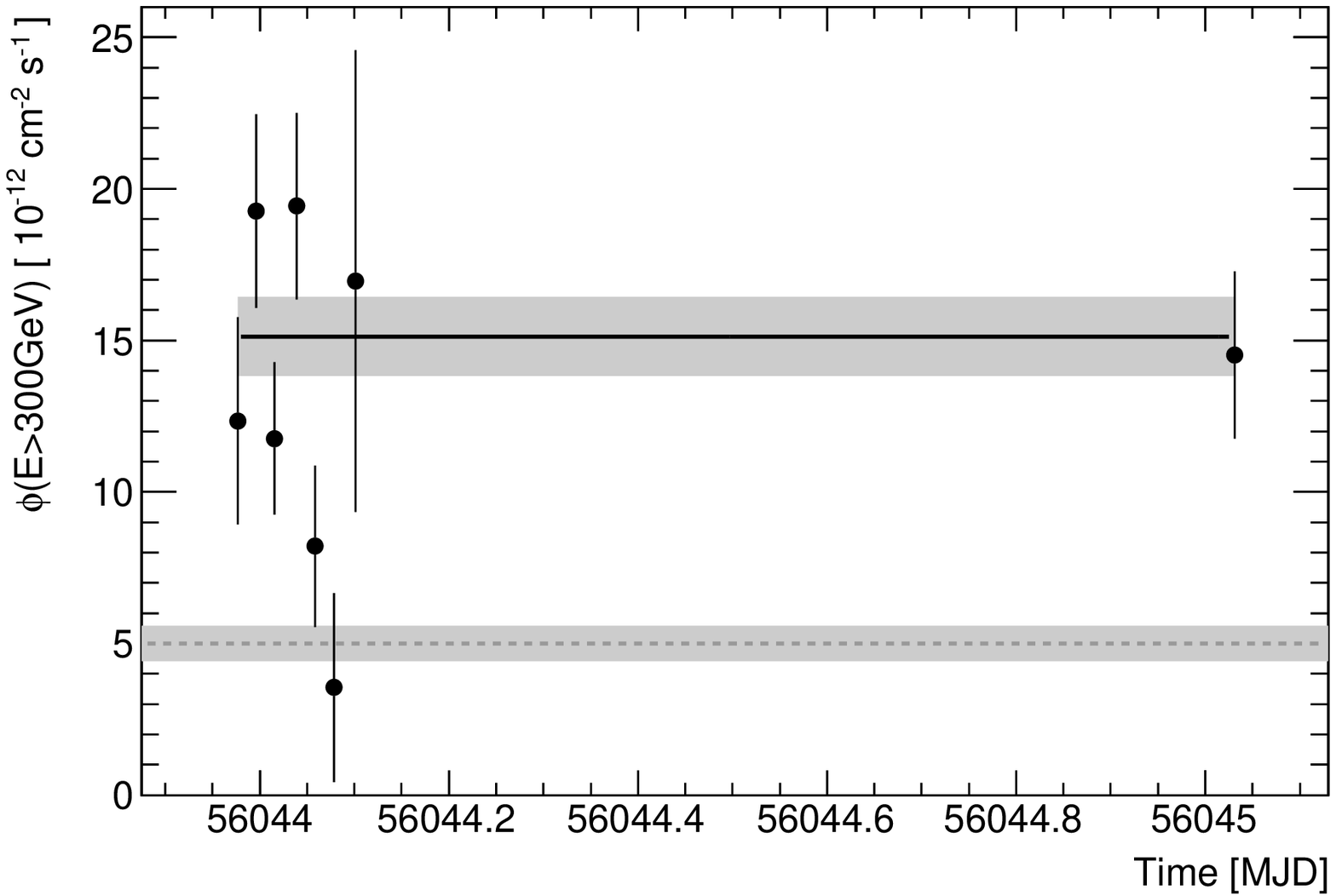}
\caption{\hess\ light curve of \pg\ during the 2 nights of the flare period. The
continuous line is the measured flux during the flare period while the dashed one
corresponds to the pre-flare period (see Table~\ref{tab:specvhe} for the flux values).
Gray areas are the $1 \, \sigma$ errors.}
\label{fig:hesslc}
\end{figure}

To compute the light curves, the integrated flux above 300 GeV for each
observation run was extracted using the corresponding (pre-flare or flare)
best fit spectral model. A fit with a constant of the run-wise light curve of the entire (pre-flare+flare)
data set, weighted by the statistical errors yields a $\chi^2$ of $123.2$ with
$68$ d.o.f. ($P_{\rm \chi^2}=6.6\times10^{-5}$). Restricting the analysis to the pre-flare data set
only, the fit yields a $\chi^2$ of 51.76 with 60 d.o.f. ($P_{\chi^2} = 0.77$),
indicating again a flux increase detected by \hess\ at the time of the flaring activity
reported by \citet{2012ATel.4069....1C}.

Figure \ref{fig:hesslc} shows the light curve during the flare together with the averaged
integral fluxes above 300 GeV of both data sets. A fit with a constant to the \hess\
light curve during the first night yields a $\chi^2$ of $20.76$ for $6$ d.o.f.
($P_{\rm \chi^2}=2.0\times 10^{-3}$), indicating intra-night variability. This is also supported by the use of a Bayesian block algorithm \citep{1998ApJ...504..405S} that finds three blocks for the 2 nights at a 95\% confidence level.

\begin{deluxetable}{lccc}
\tablecaption{Summary of the fitted spectral
      parameters for the pre-flare and the flare data sets and the 
      corresponding integral flux $I$ calculated above 300~GeV. The last column gives the 
      decorrelation energy.\label{tab:specvhe}}

\tablewidth{0pt}
\tablehead{\colhead{Data Set (Model)} & \colhead{Spectral Parameters} &\colhead{$I$ (E$>$300~GeV)}&\colhead{$E_{\rm dec}$}\\
&  &  $[10^{-12}~\rm{ph}~$\cms$]$ & [GeV]
}
\startdata
    Pre-Flare (PWL)        & $\Gamma~=~4.8\pm0.2_{\rm{stat}}\pm0.2_{\rm{sys}}$ &  $4.4\pm0.4_{\rm{stat}}\pm0.9_{\rm{sys}}$ & 306 \\ \hline
    Pre-Flare (LP) & $a~=~5.4\pm0.4_{\rm{stat}}\pm0.1_{\rm{sys}}$ &  $5.0\pm0.6_{\rm{stat}}\pm1.0_{\rm{sys}}$&\nodata  \\
    ~ & $b~=~4.0\pm1.4_{\rm{stat}}\pm0.2_{\rm{sys}}$ & ~  \\ \hline\hline
    Flare (PWL)        & $\Gamma~=~4.9\pm0.3_{\rm{stat}}\pm0.2_{\rm{sys}}$  &  $15.1\pm1.3_{\rm{stat}}\pm3.0_{\rm{sys}}$ & 327 \\ 
\enddata
\end{deluxetable}

\subsection{\fla\ analysis}\label{anafermi}
The \fer\ Large Area Telescope (LAT) is detector converting $\gamma$ ray to $e^+e^-$ pairs 
\citep{2009ApJ...697.1071A}. The LAT is
sensitive to $\gamma$ rays from 20~MeV to $>300$~GeV. In survey mode, in which the bulk of
the observations are performed, each source is seen every 3 hours for
approximately 30 minutes.

The \fla\ data and software are available from the \fer\ Science Support
Center\footnote{\url{http://fermi.gsfc.nasa.gov/ssc/data/analysis/}}. In this
work, the ScienceTools {\tt V9R32P5} were used with the Pass~7 reprocessed data
\citep{2013arXiv1304.5456B}, specifically {\tt SOURCE} class event
\citep{2012ApJS..203....4A}, with the associated {\tt P7REP\_SOURCE\_V15}
instrument response functions (IRFs). Events with energies from 300~MeV to
300~GeV were selected. Additional cuts on the zenith angle ($<100^{\circ}$) and
rocking angle ($<52^{\circ}$) were applied as recommended by the LAT
collaboration\footnote{\url{http://fermi.gsfc.nasa.gov/ssc/data/analysis/documentation/Cicerone/index.html}}
to reduce the contamination from the Earth atmospheric secondary radiation.

The analysis of the LAT data was performed using the {\tt Enrico} Python package
\citep{2013arXiv1307.4534S}. The sky model was defined as a region of interest
(ROI) of $15^\circ$ radius with \pg\ in the center and additional point-like
sources from the internal 4-years source list. Only the sources within a 3\dgr\
radius around \pg\ and bright sources (integral flux greater that $5\times10^{-7}\,$ph\,\cms) had their parameters free to vary during
the likelihood minimization. The template files 
{\tt isotrop\_4years\_P7\_V15\_repro\_v2\_source.txt} for the
isotropic diffuse component, and {\tt
template\_4years\_P7\_v15\_repro\_v2.fits} for the standard Galactic model, were
included. A binned likelihood analysis \citep{1996ApJ...461..396M}, implemented
in the {\tt gtlike} tool, was used to find the best-fit parameters. 

\begin{figure}[ht]
		 \centering
     \includegraphics[width=0.9\textwidth]{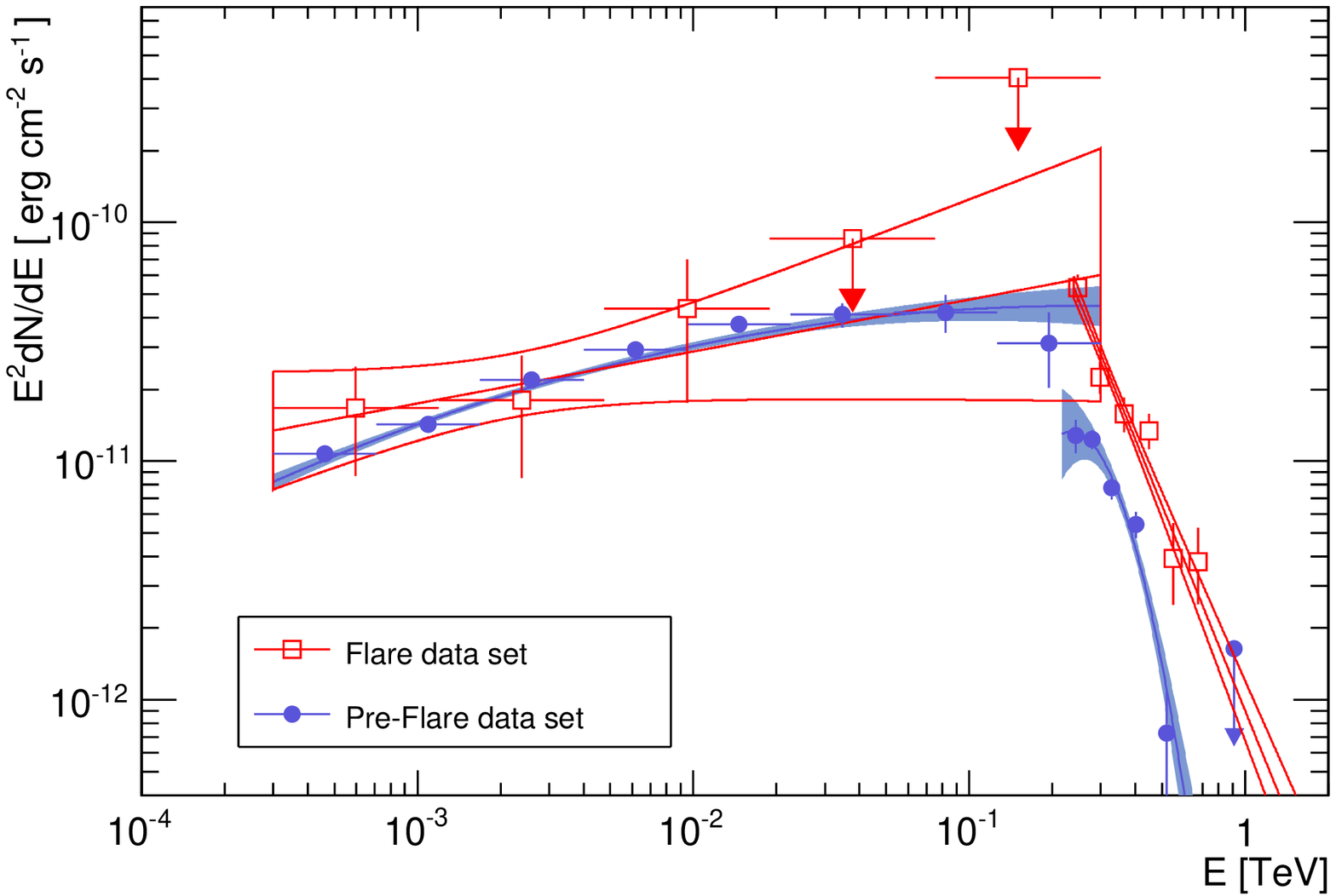}
\caption{Spectral energy distribution of \pg\ in $\gamma$ rays as measured by the \fla\ and H.E.S.S. Red (blue) 
points and butterflies have been obtained during the flare (pre-flare) period. The \textit{Fermi} and H.E.S.S. data for the pre-flare are not contemporaneous. H.E.S.S. data were taken in 2005-2006 while the \textit{Fermi} data were taken between 2008 and 2012.}
\label{fig:HESED}
\end{figure}

As for the \hess\ data analysis, two spectral models were used: a simple PWL and
a LP. A likelihood ratio test was used to decide which model best describes the
data. Table~\ref{fermiRes} gives the results for the two time periods considered
in this work, and Figure \ref{fig:HESED} presents the \gr\ SEDs. The first one
(pre-flare), before the \hess\ exposures in 2012, includes more than 3.5 years
of data (from 2008 August 4 to 2012 March 1). The best fit model is found to be
the LP (with a TS\footnote{Here the TS is 2 times the difference between the
log-likelihood of the fit with a LP minus the log-likelihood with a PL.} of
11.3, $\approx 3.4\sigma$). The second period (flare) is centered on the \hess\
observations windows and lasts for seven days. The best fit model is a power
law, the flux being consistent with the one measured during the first 3.5 years.
Data points or light curves were computed within a restricted energy range or
time range using a PWL model with the spectral index frozen to 1.70. 

To precisely probe the variability in HE $\gamma$ rays, seven-day time bins were 
used to compute the light curve of \pg\ in an extended time window (from 2008
August 4 to 2012 October 30), to probe any possible delay of a HE flare with 
respect to the VHE one. While the flux of \pg\ above 300~MeV is found to be variable 
in the whole period with a variability index of $\fvar=0.16 \pm 0.04$ \citep{Vaughan}, 
there is no sign of any flaring activity
around the 2012 \hess\ observations. This result has been confirmed by using the
Bayesian block algorithm, which finds no block
around the \hess\ exposures in 2012. Similar results were obtained when
considering only photons with an energy greater than 1~GeV. No sign of
enhancement of the HE flux associated to the VHE event reported here was found. This might be due to the lack of statistic at high energy in the LAT energy range.


\begin{deluxetable}{lcllc}
\tablecaption{Results of the \fla\ data analysis for the pre-flare and flare periods. 
For the latter, the analysis has been performed in two energy ranges (see \ref{ebl}). 
The first columns give the time and energy windows 
and the third the corresponding test statistic (TS) value. The model 
parameters and the flux above 300~MeV are given in the last columns. The systematic uncertainties were computed 
using the IRFs bracketing method \citep{2009ApJ...707.1310A}.\label{fermiRes}}
\tablewidth{0pt}
\tablehead{\colhead{MJD range} &\colhead{Energy range} & \colhead{TS} &\colhead{Spectral Parameters} &\colhead{$I$(E$>300$MeV)} \\
 & [GeV] & & & $10^{-8}$[ph \cms] 
}
\startdata

54682-55987 & 0.3-300 &7793.7 &  $a~=~1.49\pm0.06_{\rm stat}\pm 0.01_{\rm sys}$ &$2.82\pm0.1_{\rm stat}\pm 0.2_{\rm sys}$  \\
 &  & &$b~=~3.8\pm1.1_{\rm stat}\pm 0.1_{\rm sys}$ &  \\
56040-56047& 0.3-300  &43.8 & $\Gamma~=~1.78\pm0.24_{\rm stat}\pm 0.01_{\rm sys}$&$3.5\pm1.3_{\rm stat}\pm 0.3_{\rm sys}$  \\\hline
56040-56047&0.3-80 & 44.5 & $\Gamma~=~1.72\pm0.26_{\rm stat}\pm 0.01_{\rm sys}$ &  $3.4\pm1.3_{\rm stat}\pm 0.3_{\rm sys}$   \\

\enddata
\end{deluxetable}

\section{Discussion of the results}\label{discussion}

\subsection{Variability in \grs}

The VHE data do not show any sign of variation of the spectral index (when
comparing flare and pre-flare data sets with the same spectral model), and in HE no counterpart of this event can be found. The indication for intra-night
variability is similar to other TeV HBLs (Mrk~421, Mrk~501 or PKS~2155-304) with,
in this case, flux variations of a factor~3.

As noticed in previous works, \pg\ presents a sharp break between the HE and VHE
ranges \citep{2010ApJ...708.1310A} and the peak position of the $\gamma$-ray 
spectrum in the $\nu f(\nu)$
representation is located around 100 GeV. This is confirmed by the fact that the
log-parabola model better represents the pre-flare period in HE. Nonetheless,
the precise location of this peak cannot be determined with the \fla\ data only.
Combining both energy ranges and fitting the HE and VHE data points with a power
law with an exponential cutoff\footnote{A fit with a LP model has been attempted, but 
the power law with an exponential cutoff leads to a better description of the data.} 
allows us to determine the $\nu f(\nu)$ peak position for both time periods. The functional form of the model is $$ E^2 \frac{dN}{dE} = \mathrm{N} \left(\frac{E}{\mathrm{100~GeV}}\right)^{-\Gamma}\exp(-E/E_{\rm c}).$$ 

For this 
purpose, \fla\ and \hess\ systematic uncertainties were taken into account in a 
similar way as in \citet{aplib} and added quadratically to the statistical errors. 
The \fla\ systematic uncertainties
were estimated by \citet{2012ApJS..203....4A} to be 10~\% of the effective area at
100~MeV, 5~\% at 316~MeV and 15~\% at 1~TeV and above. For
the VHE $\gamma$-ray range, they were taken into account by shifting
the energy by 10~\%. This effect translates into a systematic
uncertainty for a single point of $\sigma({\rm
f})_{\rm sys}=0.1\cdot \partial{\rm f}/\partial E$ where $f$ is the differential flux at energy $E$.

The results of this parameterization are given in Table~\ref{fit}. Using the
pre-flare period, the peak position is found to be located at $\log_{10}(E_{\rm
max}/1~{\rm GeV})=1.7\pm0.2_{\rm stat}\pm0.4_{\rm sys}$ with no evidence of
variation during the flare and no spectral variation. This is
consistent with the fact that no variability in HE $\gamma$ rays was found
during the \hess\ observations. This is also in agreement with the fact that
HBLs are less variable in HE $\gamma$ rays than other BL Lac objects
\citep{2010ApJ...722..520A}, while numerous flares have been reported in the TeV
band. 

\begin{deluxetable}{cccccc}
\tablecaption{Parametrization results of the two time periods (first
column) obtained by combining \hess\ and \fla. The second column gives the 
normalization at 100~GeV, while the third
and the fourth present the spectral index and cut-off energy of the fitted power
law with an exponential cut-off. The last column is the peak energy in a $\nu
f(\nu)$ representation.\label{fit}}
\rotate
\tablewidth{0pt}
\tablehead{\colhead{Period} & \colhead{N ($E$=100~GeV)} &\colhead{$\Gamma$ }  
&\colhead{$\log_{10}(E_{\rm c}/1~{\rm GeV})$}&\colhead{$\log_{10}(E_{\rm max}/1~{\rm GeV})$}\\
 &    $10^{-11}$[\ergcms] &&  &&
}
\startdata
Pre-Flare & $9.6\pm0.7_{\rm stat}\pm1.7_{\rm sys}$ & $1.59\pm0.02_{\rm stat}\pm0.03_{\rm sys}$ & $2.03\pm0.02_{\rm stat}\pm0.04_{\rm sys}$ & $1.7\pm0.2_{\rm stat}\pm0.4_{\rm sys}$ \\
Flare & $13.0\pm3.5_{\rm stat}\pm5.7_{\rm sys}$ & $1.56\pm0.08_{\rm stat}\pm0.11_{\rm sys}$ & $2.16\pm0.04_{\rm stat}\pm0.09_{\rm sys}$ & $1.8\pm0.7_{\rm stat}\pm1.3_{\rm sys}$ \\\hline
\enddata

\end{deluxetable}

\subsection{Constraints on the redshift}\label{ebl}

The extragalactic background light (EBL) is a field of UV to far infrared
photons produced by the thermal emission from stars and reprocessed starlight 
by dust in galaxies \citep[see][for a review]{EBL_REVIEW} that interacts with 
very high energy $\gamma$ rays from sources at cosmological distances. As a consequence, a
source at redshift $z$ exhibits an observed spectrum $\phi_{obs}(E)=\phi_{\rm
int}(E)\times e^{- \tau(E,z)}$ where $\phi_{\rm int}(E)$ is the intrinsic source
spectrum and $\tau$ is the optical depth due to interaction with the EBL. Since 
the optical depth increases with increasing $\gamma$-ray energy, the integral 
flux is lowered and the spectral index is 
increased\footnote{For sake of simplicity it is assumed here that the best-fit
model is a power law, an assumption which is true for most of the cases due to
limited statistics in the VHE range}. In the following, the model of
\citet{2008A&A...487..837F} was used to compute the optical depth $\tau$ as a
function of redshift and energy. In this section, the data taken by both
instruments during the flare period are used, with the \fla\ analysis restricted 
to the range 300~MeV$<E<$80~GeV (see Table~\ref{fermiRes} for the results). In the modest redshift range of VHE emitters detected so
far ($z\leq 0.6$), the EBL absorption is negligible below 80~GeV ($\tau_{\gamma\gamma}
\sim 0.1$ at 80~GeV for $z=0.6$). 

A measure of the EBL energy density was obtained by \citet{2012Sci...338.1190A} and
\citet{2012arXiv1212.3409H} based on the spectra of sources with a known $z$. In
the case of \pg, for which the redshift is unknown, the effects of the EBL on
the VHE spectrum might be used to derive constraints on its distance. Ideally,
this would be done by comparing the observed spectrum with the intrinsic one but
the latter is unknown. The \fla\ spectrum, derived below 80~GeV, can be 
considered as a proxy for the intrinsic spectrum in the VHE regime, or at least, as 
a solid upper limit (assuming no hardening of the spectrum).

Following the method used by \citet{2013A&A...552A.118H}, it has been assumed
that the intrinsic spectrum of the source in the H.E.S.S. energy range cannot be
harder than the extrapolation of the \textit{Fermi-LAT} measurement. From this,
one can conclude that the optical depth cannot be greater than $\tau_{\rm max}(E)$,
given by:

\begin{equation}
\tau_{\rm max}(E) = \ln\left[\frac{\phi_{\rm int}}{(1-\alpha)(\phi_{\rm
obs}-1.64\Delta\phi_{\rm obs})}\right],
\end{equation}
where $\phi_{\rm int}$ is the extrapolation of the \textit{Fermi-LAT}
measurement towards the H.E.S.S. energy range. $\phi_{\rm obs}\pm\Delta\phi_{\rm
obs}$ is the measured flux by H.E.S.S. The factor $(1-\alpha) = 0.8$ accounts
for the systematic uncertainties of the H.E.S.S. measurement and the number 1.64
has been calculated to have a confidence level of 95\%
\citep{2013A&A...552A.118H}. The comparison is made at the \hess\ decorrelation
energy where the flux is best measured.

\begin{figure}[ht]
 \centering
     \includegraphics[width=0.9\textwidth]{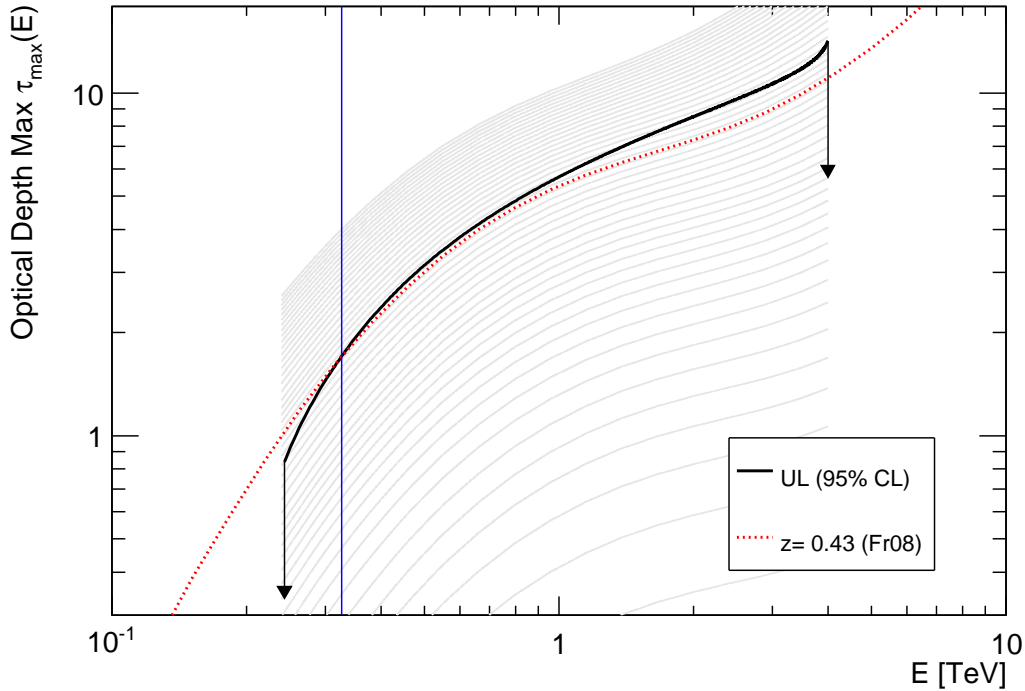}
\caption{Values of $\tau_{\rm max}$ as a function of the photon energy. The
black line is the 95\% UL obtained with the H.E.S.S. data and the red line is
the optical depth computed with the model of \cite{2008A&A...487..837F} for a
redshift of 0.43. The blue line is the decorrelation energy fo the \hess\ analyse. The gray lines are the value of optical depth for different redshift.}
\label{fig:taumax}
\end{figure}

Figure \ref{fig:taumax} shows the 95~\% UL on $\tau_{\rm max}$. The resulting
upper limit on the redshift is $z<0.43$. This method does not
allow the statistical and systematic uncertainties of the \fla\ measurement to
be taken into account and does not take advantage of the spectral features of
the absorbed spectrum \citep[see][]{2012arXiv1212.3409H}.

A Bayesian approach has been developed with the aim of taking all the
uncertainties into account. It also uses the fact that
EBL-absorbed spectra are not strictly power laws. The details of the model are
presented in Appendix \ref{BM} and only the main assumptions and results are
recalled here. Intrinsic curvature between the HE and VHE ranges that naturally
arises due to either curvature of the emitting distribution of particles or
emission effects (e.g. Klein-Nishina effects) is permitted by construction of
the prior (Eq.~\ref{eq:prior}): A spectral index softer than the \fla\
measurement is allowed with a constant probability, in contrast with the
previous calculation. It is assumed that the observed spectrum in VHE
$\gamma$ rays cannot be harder than the \fla\ measurement by using a prior that
follows a Gaussian for indices harder than the \fla\ one. The prior on the index is then:

\begin{equation} 
P(\Gamma) \propto \mathcal{N}_G (\Gamma,\Gamma_{\rm Fermi},\sigma_\Gamma)
\end{equation}
if $\Gamma<\Gamma_{\rm Fermi}$ and $$P(\Gamma) \propto 1 $$ otherwise.
$\Gamma_{\rm Fermi}$ is the index measured by \fla\ and $\sigma_\Gamma$ is the
uncertainty on this measurement that takes all the systematic and statistical
uncertainties into account.

\begin{figure}[ht]
		 \centering
     \includegraphics[width=0.9\textwidth]{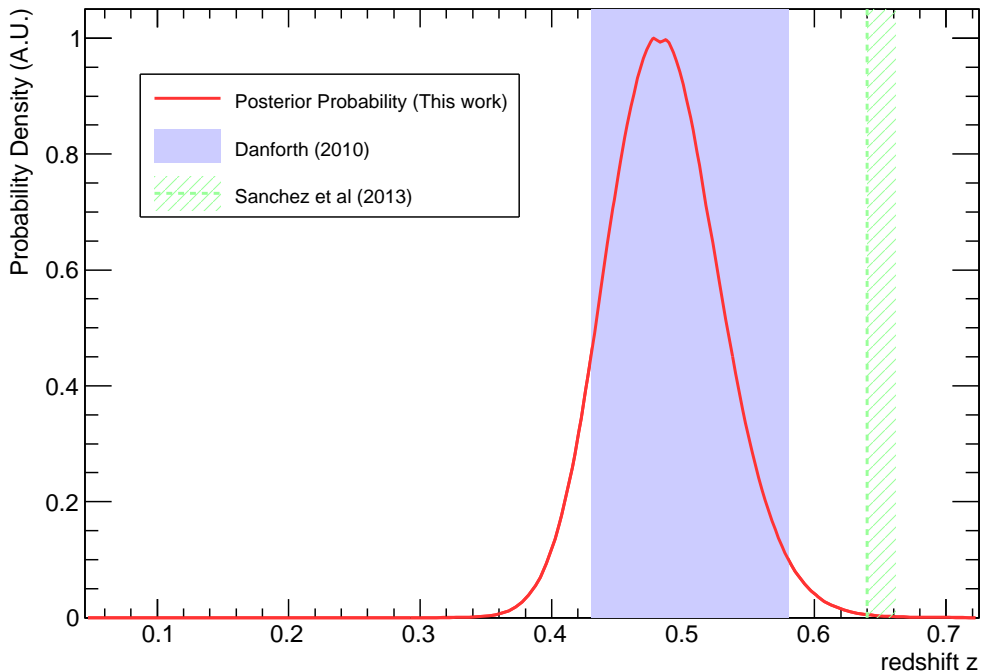}
\caption{Posterior probability density as a function of redshift (red). The blue area 
represents the redshift range estimated by \citet{danforth_hubble/cos_2010} while the 
green dashed line indicates the limit of \citet{2013A&A...554A..75S}.}
\label{fig:Prob}
\end{figure}

The most probable redshift found with this method is \bestz, in good agreement
with the independent measure of \citet{danforth_hubble/cos_2010}, who constrained 
the distance to be between $0.43<z<0.58$. Figure \ref{fig:Prob} gives the
posterior probability obtained with the Bayesian method compared with other
measurements of $z$. Lower and upper limits at a confidence level of 95~\% can
also be derived as $0.41<z<0.56$. Note that this method allows the systematic
uncertainties of both instruments (\fla\ and \hess) to be taken into account. The spectral index obtained when fitting the H.E.S.S. data with an EBL absorbed PWL using a redshift of 0.49 is compatible with the \textit{Fermi} measurement below 80 GeV.

\subsection{Lorentz Invariance Violation}
As stated in section~\ref{anahess}, the \hess\ data of the flare show a indication of intra-night variability, which is used here to test for a possible Lorentz Invariance Violation (LIV). Some Quantum Gravity (QG) models predict a change
of the speed of light at energies close to the Planck scale ($\sim
10^{19}$\,GeV). A review of such models can be found in
\cite{mattingly_modern_2005} and \cite{liberati_tests_2013}. An energy-dependent
dispersion in vacuum is searched for in the data by testing a correlation
between arrival times of the photons and their energies. For two photons with
arrival times $t_1$ and $t_2$ and energies $E_1$ and $E_2$, the dispersion
parameter of order $n$ is defined as \mbox{$\tau_n = \frac{t_2-t_1}{E_2^n -
E_1^n} = \frac{\Delta t}{\Delta (E^n)}$}. Here only the linear \mbox{(n = 1)}
and quadratic \mbox{(n = 2)} dispersion parameters are calculated. Assuming no
intrinsic spectral variability of the source, the dispersion $\tau_n$ can be
related to the normalized distance of the source $\kappa_n$ corrected for the 
expansion of the Universe and an energy $\textrm{E}_{\rm QG}$ at which Quantum Gravity 
effects are expected to occur \citep{jacob_lorentz_violation_induced_2008}:
\newline
\begin{equation}
\label{eq:tau}
\tau_n = \frac{\Delta t}{\Delta (E^n)} \simeq s_{\pm}\frac{(1+n)}{\textrm{E}_{\rm QG}^n 2\mathrm{H}_\mathrm{0}} 
\kappa_n
\end{equation}
where H$_{0}$ is the Hubble constant and \mbox{$s_{\pm}$ = $-$1} (resp. +1) in
the superluminal (resp. subluminal) case, in which the high-energy photons 
arrive before (resp. after) low-energy photons. The normalized distance $\kappa_n$ is
calculated from the redshift of the source $z$ and the cosmological parameters
$\Omega_m$, $\Omega_\Lambda$ given in the introduction: 
\newline
\begin{equation}
\kappa_n = \int_0^z \frac{(1+z')^n\,dz'}{\sqrt{\Omega_m(1+z')^3 + \Omega_{\Lambda}}}
\end{equation}
Using the central value of $z=0.49$ determined in section \ref{ebl}, the distance $\kappa_n$ f
or n = 1 and 2 is $\kappa_1 = 0.541$ and $\kappa_2 = 0.677$.

First, the dispersion measurement method will be described. It will then be applied to 
the \hess\ flare dataset (MC simulations and original dataset), in order to measure the 
dispersion and provide 95~\% 1-sided lower and upper limits on the dispersion parameter 
$\tau_n$. These limits on $\tau_n$ will lead to lower limits on $\textrm{E}_{\rm QG}$ using 
equation \ref{eq:tau}.

\subsubsection{Modified maximum likelihood method}

A maximum likelihood method, following \cite{martinez_new_2009}, was used to 
calculate the dispersion parameter $\tau_n$. \citet{albert_probing_2008} applied
this method to a flare of Mkn~501, while \citet{abramowski_search_2011}
applied it to a flare of \mbox{PKS~2155-304}. More recently, it was used by 
\cite{vasileiou_constraints_2013} to analyse \textit{Fermi} data of four gamma-ray bursts. 
The data from Cherenkov telescopes is contaminated by $\pi^0$ decay from proton showers, 
misidentified electrons, or heavy
elements such as helium. In the case of \pg, and contrary to previous analyses, this 
background is not negligible: the signal-over-background ratio S/B is about 2, compared 
to 300 for the PKS\,2155--304 flare event of July 2006 \citep{2007ApJ...664L..71A}. 
The background was included in the formulation of the probability density function 
(PDF) used in a likelihood maximization method. 
Given the times $t_i$ and energies $E_i$ of the gamma-like (ON) particles received 
by the detector, the unbinned likelihood, function of the dispersion parameter $\tau_n$ is:
\begin{equation}
\newline
\mathit{L}(\tau_n) = \prod_{i = 1}^{n_{\rm ON}} P(E_i, t_i|\tau_n).
\newline
\label{eqliketext}
\end{equation}
The PDF $P(E_i, t_i|\tau_n)$ associated with each ON
event is composed of two terms:  
\begin{align}
\newline
&P(E_i, t_i|\tau_n)  
= w_s \cdot P_{\textrm{\tiny Sig}}(E_i, t_i|\tau_n) + (1-w_s) \cdot P_{\textrm{\tiny Bkg}}(E_i, t_i) 
\end{align}
with
\begin{align}
P_{\textrm{\tiny Sig}}(E_i, t_i|\tau_n) &= \frac{1}{N(\tau_n)} \, A_{\rm eff}(E_i,t_i) \, 
\Lambda_{\textrm{\tiny Sig}}(E_i) \, F_{\textrm{\tiny Sig}}(t_i - \tau_n \cdot E_i^n) \\
P_{\textrm{\tiny Bkg}}(E_i, t_i) &= \frac{1}{N'} \, A_{\rm eff}(E_i,t_i) \, 
\Lambda_{\textrm{\tiny Bkg}}(E_i) \, F_{\textrm{\tiny Bkg}}\\
w_s &= \frac{n_{\rm ON}-\alpha \, n_{\rm OFF}}{n_{\rm ON}}.
\newline
\end{align}
The PDF $P_{\textrm{\tiny Sig}}$ includes the emission time distribution of the photons 
$F_{\textrm{\tiny Sig}}$ determined from a parametrization of the observed light curve 
at low energies (discussed in the next section) and evaluated on $t - \tau_n \cdot E^n$ 
to take into account the delay due to a possible LIV effect, the measured signal spectrum 
$\Lambda_{\textrm{\tiny Sig}}$ and the effective area $A_{\rm eff}$.
The PDF $P_{\textrm{\tiny Bkg}}$ is composed of the uniform time distribution $F_{\textrm{\tiny Bkg}}$ 
of the background events, the measured background spectrum $\Lambda_{\textrm{\tiny Bkg}}$ and the 
effective area $A_{\rm eff}$. No delay due to a possible LIV effect is expected in the background 
events of the ON data set.
$N(\tau_n)$ and $N'$ are the normalization factors of $P_{\textrm{\tiny Sig}}$ and 
$P_{\textrm{\tiny Bkg}}$ respectively, in the ($E$, $t$) range of the likelihood fit.
The coefficient $w_s$ corresponds to the relative weight of the signal events in the 
total ON data set, derived from the number of events in the ON region $n_{\rm ON}$ and 
the number of events in the OFF regions $n_{\rm OFF}$ weighted by the inverse number of 
OFF regions $\alpha$. More details on the derivation of this function are given in 
Appendix \ref{app:like}.

\subsubsection{Specific selection cuts and timing model}

\begin{figure}
\centering
\includegraphics[scale=0.6]{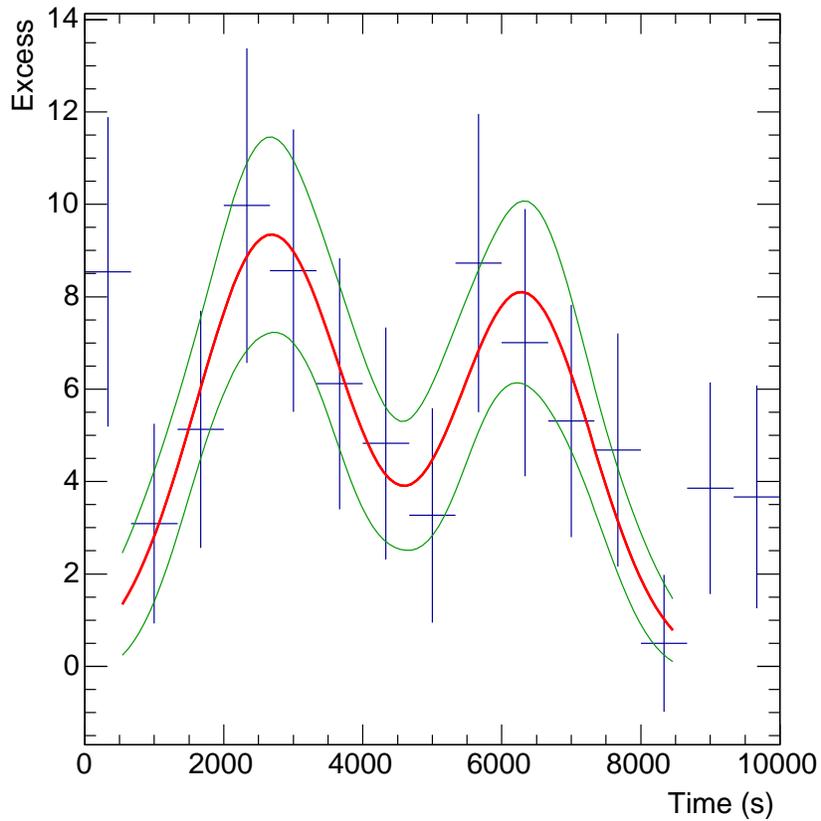}
\caption{Time distribution of the excess $ON - \alpha OFF$ in the first 6 runs (70971-70976), 
with energies between 300\,GeV and 400\,GeV.
T = 0 corresponds to the time of the first detected event in run 70971. 
The vertical bars correspond to $1 \, \sigma$ statistical errors; the horizontal bars 
correspond to the bin width in time. The best fit, in red, was used as the template light 
curve in the maximum likelihood method; the $\pm 1\,\sigma$ error envelope is shown in green.}
\label{fig:template}
\end{figure}

The flare data set of the H.E.S.S. analysis (see section \ref{anahess}) was used with 
additional cuts. To perform the dispersion studies, only uninterrupted data have been 
kept. Thus, the analysis was conducted on the first 7 runs, taken during the night of 
April 26th. Moreover, the cosmic ray flux increases substantially for the 7th run, due 
to a variation of the zenith angle during this night. This fact, along with its large 
statistical errors, leads us to discard this run from the analysis. The 6th run 
shows little to no variability and was therefore also removed from the LIV analysis.
Since within the ON data set, the signal and the background spectra have different
indices ($\Gamma_{\rm \tiny Sig} = 4.8$ for the signal and $\Gamma_{\rm \tiny Bkg} = 2.5$ 
for the background),
the ratio S/B is expected to decrease with increasing energy. An upper
energy cut at $E_{\rm max}$ = 789 GeV was set, corresponding to the last
bin with more than $3 \, \sigma$ significance in the reconstructed photon spectrum (see
the differential flux during the flare in Fig. \ref{fig:specvhe}). 
A lower cut on the energy at $E_{\rm min}$ = 300~GeV was used in order to avoid large 
systematic effects arising from high uncertainties on the H.E.S.S. effective area at lower energies.
The intrinsic light curve of the flare, needed in the formulation of the likelihood, can be 
obtained from a model of the timed emission or approximated from a subset of the data. To be 
as model-independent as possible, it was here derived from a fit of the measured light curve 
at low energies (with $E < E_{\rm cut}$). 
The high-energy events ($E > E_{\rm cut}$) were processed in the calculation of the likelihood 
to search for potential dispersion. Here $E_{\rm cut}$ was set to $E_{\rm cut}$ = 400~GeV, 
which is approximately the median energy of the ON event sample. Other cuts on the energy 
did not introduce significant effects on the final results. 
The histogram and the fit (Fig.~\ref{fig:template}) were obtained as follows: the main idea was to preserve the 
maximum detected variability in the \pg~flare, together with a significant response in 
each observed peak:
\begin{itemize}
\item The binning was chosen so that at least two adjacent bins of the 
distribution yield a minimum of $3 \, \sigma$ excess with respect to the average value. 
\item Simple parameterization have been tested on the whole data set (all
energies): constant ($\chi^2$/d.o.f=25/12), single Gaussian
($\chi^2$/d.o.f=20/10) and double Gaussian ($\chi^2$/d.o.f=8.5/7) functions.
The latter is preferred, since it improves the quality of the fit. This shape
was chosen to fit the low energy subset of events. Choosing a single Gaussian
parametrization would result in a decrease of the sensitivity to time-lag
measurements by a factor of two.
\end{itemize}

There is a gap of $\sim$\,2\,min between each two consecutive runs. We did not
consider the effect of these gaps as it is small with respect to the bin width
of $\sim$\,10\,min. More importantly, their occurrence is not correlated with
the binning: one gap falls in the rising part of the light curve, one is at a
maximum, two fall in the decreasing parts and none of the gaps is at the
minimum.

Table \ref{tab:cuts} in Appendix \ref{app:selcuts} shows the number of ON and OFF events 
for the different cuts applied to the data. 

\subsubsection{Results: limits on $\tau_n$ and $\textrm{E}_{\rm QG}$}

The maximum likelihood method was performed using high-energy events with \mbox{$E_i>E_{\rm cut}$}.
First, confidence intervals (CIs) corresponding to 95~\% confidence level (1-sided)
were determined from the likelihood curve at the values of $\tau_n$ where the curve
reaches 2.71, which corresponds to the 90\% C.L. quantile of a $\chi^2$ distribution. 
However, these CIs are derived from one realization only and do not take into account 
the ``luckiness'' factor of this measurement. To get statistically
significant CIs (``calibrated CIs''), several sets were generated with Monte Carlo 
simulations, with the same statistical significance, light curve model and
spectrum as the original data set. No intrinsic dispersion was artificially
added. Each simulated data set produces a lower limit and an upper limit on
$\tau_n$. The calibrated lower (upper) limit of the confidence interval is
obtained from the mean of the distribution of the per-set individual lower (upper)
limits.
Both confidence intervals (from the data only and from the simulated sets) are 
listed in Table
\ref{tab:llultau}. Sources of systematic errors include uncertainties on the light 
curve parameterization, the background contribution, the calculation of the effective 
area, the energy resolution, and the determination of the photon index (see Appendix 
\ref{annexe:syst}).

\begin{deluxetable}{lcccc}
\tablecaption{Calibrated 95\% 1-sided LL and UL (including systematic errors) on 
the dispersion parameter $\tau_n$ and derived 95\% 1-sided lower limits on $\textrm{E}_{\rm QG}$. 
\label{tab:taueqg}}
\tablewidth{0pt}
\tablehead{ &  \multicolumn{2}{c}{Limits on $\tau_n$ (s\,TeV$^{-n}$)}    &  
\multicolumn{2}{c}{Lower limits on $\textrm{E}_{\rm QG}$ (GeV)}
\\
n & $LL^{\rm calib+syst}$ & $UL^{\rm calib+syst}$ &     s$= -1$     &     s$= +1$   
}
\startdata

	1 &      -838.9       &       576.4       & 2.83 10$^{17}$ & 4.11 10$^{17}$ \\ 
	2 &      -1570.5      &       1012.4      & 1.68 10$^{10}$ & 2.10 10$^{10}$ \\ 
\enddata
\end{deluxetable}

\begin{figure}
\centering
\includegraphics[scale=0.73]{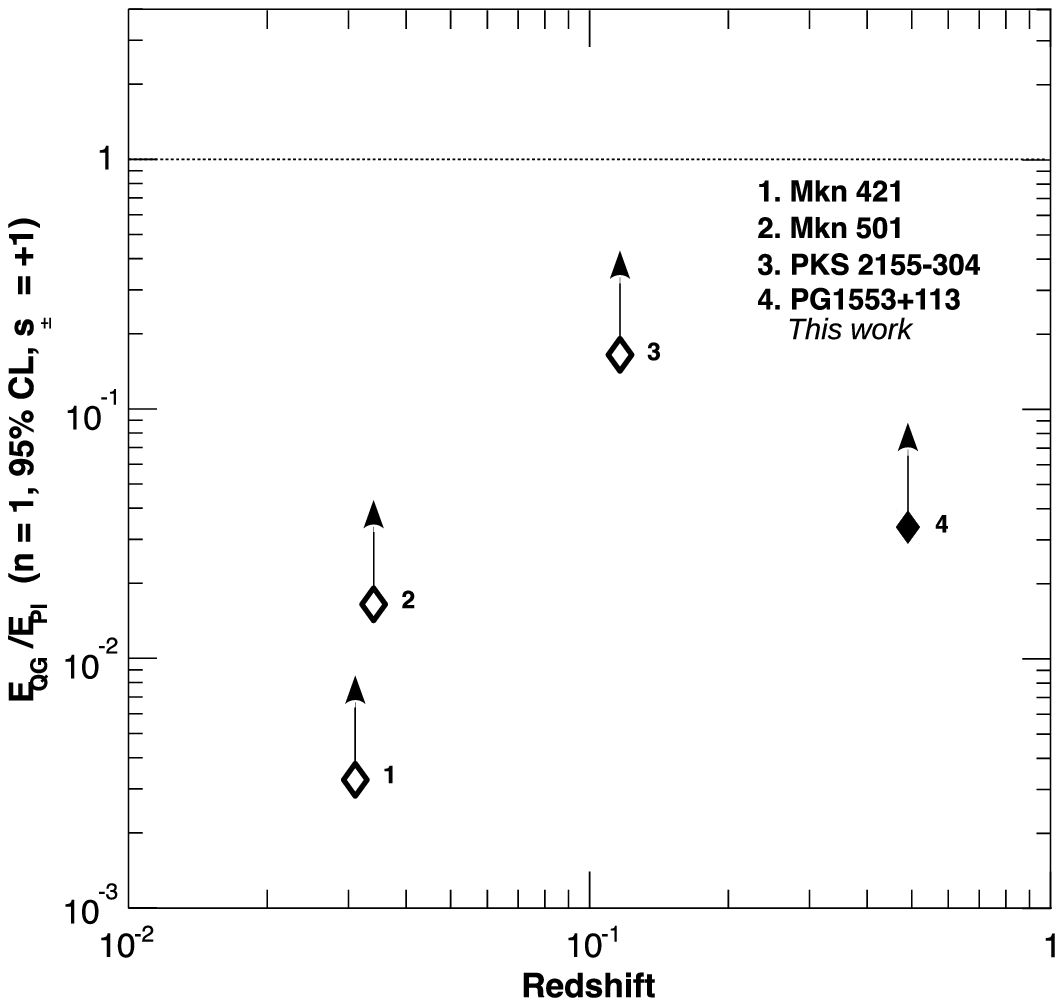}
\includegraphics[scale=0.73]{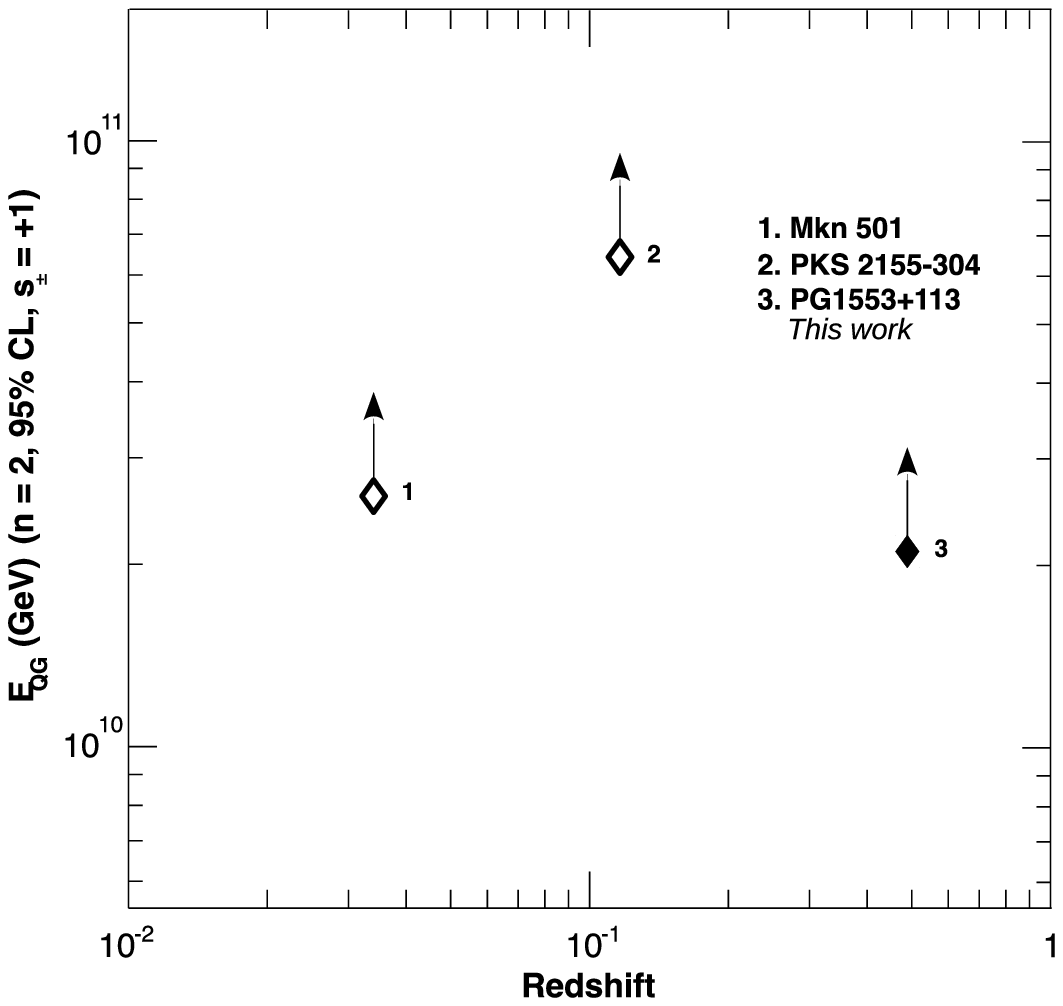}
\caption{Lower limits on $\textrm{E}_{\rm QG,1}$ from linear dispersion (left) 
and on $\textrm{E}_{\rm QG,2}$ from quadratic dispersion (right) for the subluminal 
case (s = +1) obtained with 
AGN as a function of redshift. The limits are given in terms of $\textrm{E}_{\rm Planck}$. 
The constraints from Mkn~421 have been obtained by \citet{biller_limits_1999}, from 
Mkn~501 by \citet{albert_probing_2008}, 
and from PKS~2155-304 by \citet{abramowski_search_2011}.}
\label{fig:comparison}
\end{figure}

The resulting limits on the dispersion $\tau_n$ using the quadratic sum of
the statistical errors from the simulations and the systematic errors 
determined from data and simulations were computed,
leading to limits on the energy scale $\textrm{E}_{\rm QG}$ (Eq.~\ref{eq:tau}).
The 95~\% 1-sided lower limits for the subluminal case (s = +1) are:
$\textrm{E}_{\rm QG,1}>4.11\times 10^{17}$~GeV and $\textrm{E}_{\rm
QG,2}>2.10\times 10^{10}$~GeV for linear and quadratic LIV effects,
respectively. For the superluminal case (s = --1) the limits are: $\textrm{E}_{\rm
QG,1}>2.83\times 10^{17}$~GeV and $\textrm{E}_{\rm QG,2}>1.68\times 10^{10}$~GeV
for linear and quadratic LIV effects, respectively. 
Fig.~\ref{fig:comparison} shows a comparison of the different lower limits on 
$\textrm{E}_{\rm QG,1}$ and $\textrm{E}_{\rm QG,2}$ for the subluminal case (s = +1) 
obtained with AGN at different redshifts studied at very high energies. 
All these limits, including the present results, have been obtained under the 
assumption that no intrinsic delays between photons of different energies occur 
at the source. For the linear/subluminal case, the most constraining limit on $\textrm{E}_{\rm QG}$ with transient astrophysical events has been obtained with GRB\,090510: $\textrm{E}_{\rm QG,1} > 6.3\times10^{19}$~GeV \citep{vasileiou_constraints_2013}. The most constraining limits on $\textrm{E}_{\rm QG}$ with AGN so far have
been obtained by \cite{abramowski_search_2011} with PKS~2155-304 data observed
with H.E.S.S.: \mbox{$\textrm{E}_{\rm QG,1}>2.1\times 10^{18}$~GeV} and
\mbox{$\textrm{E}_{\rm QG,2}>6.4\times 10^{10}$~GeV} for linear and
quadratic LIV effects, respectively (95\% CL, 1-sided). 
Compared to the PKS~2155-304 limits, the limits on the linear dispersion for \pg\
are one order of magnitude less constraining, but the limits on the quadratic 
dispersion are of the same order of magnitude since the source is located at a higher 
redshift. This highlights the interest in studying distant AGN, in spite of the 
difficulties due to limited photon statistics. 

\section{Conclusions}\label{conclusion}
A VHE $\gamma$-ray flaring event of \pg\ has been detected with the \hess\ telescopes, 
with a flux increasing of a factor of 3. No variability of the spectral index has been 
found in the data set, but indication of intra-night flux variability is reported in this 
work. In HE $\gamma$ rays, no counterpart of this event can be identified, which may be 
interpreted as the sign of injection of high energy particles emitting predominantly in 
VHE $\gamma$ rays. Such particles might not be numerous enough to have a significant 
impact on the HE flux during either their acceleration or cooling phases.

The data were used to constrain the redshift of the source using a new approach based 
on the absorption properties of the EBL imprinted in the spectrum of a distant source. 
Taking into account all the instrumental systematic uncertainties, the redshift of \pg\ 
is determined as being \bestz.

Flares of variable sources can be used to probe LIV effects,
manifesting themselves as an energy-dependent delay in the photon arrival time. A
likelihood method, adapted to flares with a large amount of background and
modest statistics, was presented. To demonstrate the analysis power of this
method, it was applied to the \hess\ data of a flare of \pg. This analysis
relies on the indication of the intra-night variability of the flare at VHE.
No significant dispersion was measured, and limits on the $\textrm{E}_{\rm QG}$
scale were derived, in a region of redshift unexplored until now. 
 Limits on the energy scale at which 
QG effects causing LIV may arise, derived in this work, are $\textrm{E}_{\rm QG,1}>4.11
\times 10^{17}$~GeV and $\textrm{E}_{\rm QG,2}>2.10\times 10^{10}$~GeV for the subluminal 
case. Compared with previous limits obtained with the PKS~2155-304 flare of 2006 July, the 
limits for \pg\ for a linear dispersion are one order of magnitude less constraining while 
limits for a quadratic dispersion are of the same order of magnitude. With the new telescope 
placed at the center of the \hess\ array that provides an energy threshold of several tens 
of GeV, a better picture of the variability patterns of AGN flares should be obtained. 
The future Cherenkov Telescope 
Array (CTA) will increase the number of flare detections \citep{2013APh....43..215S} 
with better sensitivity, allowing for the extraction of even more constraining limits 
on the LIV effects.

\section*{Acknowledgements}

The support of the Namibian authorities and of the University of Namibia in
facilitating the construction and operation of H.E.S.S. is gratefully
acknowledged, as is the support by the German Ministry for Education and
Research (BMBF), the Max Planck Society, the French Ministry for Research, the
CNRS-IN2P3 and the Astroparticle Interdisciplinary Programme of the CNRS, the
U.K. Particle Physics and Astronomy Research Council (PPARC), the IPNP of the
Charles University, the South African Department of Science and Technology and
National Research Foundation, and by the University of Namibia. We appreciate
the excellent work of the technical support staff in Berlin, Durham, Hamburg,
Heidelberg, Palaiseau, Paris, Saclay, and in Namibia in the construction and
operation of the equipment.

The \textit{Fermi} LAT Collaboration acknowledges generous ongoing support from a number of agencies and institutes that have supported both the development and the operation of the LAT as well as scientific data analysis. These include the National Aeronautics and Space Administration and the Department of Energy in the United States, the Commissariat à l'Energie Atomique and the Centre National de la Recherche Scientifique / Institut National de Physique Nucléaire et de Physique des Particules in France, the Agenzia Spaziale Italiana and the Istituto Nazionale di Fisica Nucleare in Italy, the Ministry of Education, Culture, Sports, Science and Technology (MEXT), High Energy Accelerator Research Organization (KEK) and Japan Aerospace Exploration Agency (JAXA) in Japan, and the K. A. Wallenberg Foundation, the Swedish Research Council and the Swedish National Space Board in Sweden.

Additional support for science analysis during the operations phase is gratefully acknowledged from the Istituto Nazionale di Astrofisica in Italy and the Centre National d'Etudes Spatiales in France.

DS work is supported by the LABEX grant enigmass. The authors want to thanks F. Krauss for her useful comments.
\appendix
\section{Bayesian model used to constrain the redshift}
\label{BM}

A Bayesian approach has been used to compute the redshift value of PG~1553+113 in Section \ref{ebl}. The
advantage of such a model is that systematic uncertainties, which are important in
Cherenkov astronomy, can easily be included in the calculation. In the
following, the notation $\Theta$ for the model parameters and $Y$ for the data
set is adopted. All normalization constants are dropped in the development 
of the model, and the final probability is normalized at the end.

Bayes' Theorem, based on the conditional probability rule, allows us to write
the posterior probability $P(\Theta|Y)$ for the model parameters $\Theta$ as the
product of the likelihood $P(Y|\Theta)$ and the prior probability $P(\Theta)$:

\begin{displaymath}
P(\Theta|Y) \propto P(\Theta)  P(Y|\Theta).
\end{displaymath}

The likelihood is the quantity that is maximized during determination of the best-fit
spectrum \citep{2001AA...374..895P}. It is at this step that the \hess\ data, 
taken during the flare, were actually used. The spectrum model here is a simple 
power law corrected for the EBL absorption: 
$$ \phi = N\times (E/E_0)^{-\Gamma}\times e^{-\tau(E,z)}.$$ 
The model parameters are then $N$, $\Gamma$ and $z$.

The prior is the most difficult and most interesting part of the model. To
derive it, $N$ and $\Gamma$ are assumed to be independent from each other and
independent of the redshift. In contrast, the prior on the redshift might
depend on $N$ and $\Gamma$. Then, the prior can be simplified using the
conditional probability rule:

$$P(\Theta) = P(z|N,\Gamma)P(N)P(\Gamma)$$

As much as possible, weak assumptions should be made to write a robust prior then
often flat (i.e. $P\propto $ const) are used. Priors should also be based
on a physical meaning and not contradict the physical and observed properties of
the objects. For the purpose of this model, the prior on $N$ is assumed to be
flat and the prior on the spectral index is a truncated Gaussian
$P(\Gamma)\propto \mathcal{N}_G (\Gamma,\Gamma_{\rm Fermi},\sigma_\Gamma$) if
$\Gamma<\Gamma_{\rm Fermi}$ and $P(\Gamma)=\propto $ const otherwise. The values of
$\Gamma_{\rm Fermi}$ and $\sigma_{\Gamma}$ are
obtained by analyzing LAT data below 80~GeV (see section \ref{discussion} and
Table~\ref{fermiRes}). Here, it is assumed that the intrinsic spectrum in the
VHE range cannot be harder than the \fla\ measurement. $\sigma_\Gamma$ takes
into account the statistical and systematic uncertainties on the \fla\
measurement and also the systematic uncertainty on the \hess\ spectrum
\citep[$\sigma=0.20$, see][]{aha2006} added quadratically and
$\sigma_\Gamma=0.33$ for a mean value of $\Gamma_{\rm Fermi}=1.72$. 

The prior on $z$ is much more difficult to determine. A flat prior has no physical
motivations since the probability to detect sources at TeV energy decreases with
the redshift. The number of sources detected at TeV energy is not sufficient to
use the corresponding redshift distribution as a prior.

A prior which takes into account the EBL, can be derived assuming a
population of sources with a constant spatial density. In the small 
space element $4\pi z^2 dz$, the number of such sources scales $\propto z^2$.
For any given luminosity, their flux (which scales with the probability to detect 
them) is scaled by $z^{-2}exp(-\tau(z))$. Lacking a proper knowledge of the
intrinsic luminosity function of VHE $\gamma$-ray blazars, a reasonable
assumption on the detection probability of a blazar at any redshift is 
a scaling proportional to the flux for a given luminosity, i.e., $\propto z^{-2} \, exp(-\tau(z))$.
Putting everything together, the prior on the redshift reads $P(z|N,\Gamma) = P(z) \propto exp(-\tau(z))$

Finally, the prior we use for our analysis is:
\begin{equation} \label{eq:prior}
P(\Theta) \propto exp(-\tau(z)) \mathcal{N}_G (\Gamma,1.72,0.33)
\end{equation}
if $\Gamma<1.72$ and
$$P(\Theta) \propto exp(-\tau(z))$$ otherwise. Putting all the components of the model 
together and marginalizing over the nuisance parameters $N$ and $\Gamma$, the probability 
on the redshift can be computed numerically.  The 
obtained mean value is \bestz. At a confidence level of 95~\%, the redshift is 
between $0.41<z<0.56$.


In this work, only the model of \citet{2008A&A...487..837F} has been used. Other EBL 
models available in the literature predict slightly different absorption depths. 
This will lead to a small difference in the redshift. The use 
of a flat prior for the redshift distribution of the sources or a prior based on 
estimates of the HBLs luminosity function \citep{2014ApJ...780...73A} leads to 
changes of order of 0.01 on the resulting redshift.

\section{Development of the LIV method}

\subsection{Modified maximum likelihood method}
\label{app:like}

In previous LIV studies with AGN flares \citep[][]{albert_probing_2008,abramowski_search_2011} 
the signal was clearly dominating
over the background, whereas in the present study the signal-over-background ratio is about 2. 
The background has been included in the formulation of the probability density function (PDF): 
in the most general case, for given numbers of signal and background events $s$ and $b$ in the 
observation region (``ON'' region), for a given dispersion parameter $\tau_n$, the unbinned 
likelihood is:
\begin{equation}
\newline
\mathit{L}(n_{\rm ON}, n_{\rm OFF}|s,b,\tau_n) = {\rm Pois}(n_{\rm ON}| s+b) \cdot 
{\rm Pois}\left(n_{\rm OFF}| \frac{b}{\alpha }\right)\cdot \prod_{i = 1}^{n_{\rm ON}} 
P(E_i, t_i|s,b,\tau_n) 
\label{unbinnedlikesimple}
\newline
\end{equation}
The PDF $P(E_i, t_i|s,b,\tau_n)$ associated with each gamma-like
particle characterized by its time $t_i$ and energy $E_i$ contains two terms (signal and background):
\begin{equation}
\label{eq:pfull}
P(E_i, t_i|s,b,\tau_n) =   w_s \cdot P_{\textrm{\tiny Sig}}(E_i, t_i|\tau_n) + (1-w_s) \cdot 
P_{\textrm{\tiny Bkg}}(E_i, t_i) 
\end{equation}
with
\begin{equation}
w_s = \frac{s}{s+b}.
\end{equation}
$n_{\rm ON}$ is the number of events detected in the source ON region included in the fit 
range \mbox{$[E_{\rm cut};E_{\rm max}] \times [t_{\rm min};t_{\rm max}]$}. $n_{\rm OFF}$ 
is the number of events in the OFF regions, in the same ($E$, $t$) range; $\alpha$   is 
the inverse number of OFF regions. 	
${\rm Pois}(n_{\rm ON}| s+b)$ (${\rm Pois}( n_{\rm OFF}| b/\alpha)$) is the Poisson 
distribution with index $n_{\rm ON}$ ($n_{\rm OFF}$) and parameter $s+b$ ($b/\alpha$).
The likelihood function can be simplified by fixing $s$ and $b$ from a comparison of 
ON and OFF sets: \mbox{$s = n_{\rm ON} - \alpha n_{\rm OFF}$} and \mbox{$b = \alpha n_{\rm OFF}$}. 
In this case, the Poisson terms in Eq. \ref{eq:pfull} are equal to 1.
The probabilities $P_{\textrm{\tiny Sig}}$ and $P_{\textrm{\tiny Bkg}}$ are defined as:
\begin{align}
\newline
P_{\textrm{\tiny Sig}}(E_i, t_i|\tau_n) &= \frac{1}{N(\tau_n)} \cdot 
R_{\textrm{\tiny Sig}}(E_i, t_i|\tau_n)\\
P_{\textrm{\tiny Bkg}}(E_i, t_i) &= \frac{1}{N'} \cdot R_{\textrm{\tiny Bkg}}(E_i, t_i)
\newline
\end{align}
with
\begin{align}
\newline
R_{\textrm{\tiny Sig}}(E, t|\tau_n) &= \int_{E_{\rm true} = 0}^{\infty} D(E,E_{\rm true}) \, 
A_{\rm eff}(E_{\rm true}, t) \, \Lambda_{\textrm{\tiny Sig}}(E_{\rm true}) \, 
F_{\textrm{\tiny Sig}}(t - \tau_n \cdot E_{\rm true}^n) dE_{\rm true}\\
R_{\textrm{\tiny Bkg}}(E, t) &= \int_{E_{\rm true} = 0}^{\infty} D(E,E_{\rm true}) \, 
A_{\rm eff}(E_{\rm true}, t) \, \Lambda_{\textrm{\tiny Bkg}}(E_{\rm true}) \, 
F_{\textrm{\tiny Bkg}}(t) \, dE_{\rm true}.
\newline
\end{align}
$P_{\textrm{\tiny Sig}}(E_i, t_i|\tau_n)$ is the probability that the event
($E_i$, $t_i$) is a photon emitted at the source and detected on Earth
with a delay $\tau_{n} E^n$. It takes into account the emission 
(time distribution $F_{\textrm{\tiny Sig}}(t)$ and energy spectrum 
$\Lambda_{\textrm{\tiny Sig}}(E)$ at the source), the propagation 
(delay $\tau_n \cdot E_i^n$ due to possible LIV effect) and the detection of a 
photon by the detector (H.E.S.S. energy resolution $D(E,E_{\rm true})$ and effective 
area $A_{\rm eff}(E, t)$).
$P_{\textrm{\tiny Bkg}}(E_i, t_i)$ is the probability that the event ($E_i$, $t_i$) is 
a background event; it is not expected to be variable with time, thus $F_{\textrm{\tiny Bkg}}(t)$ 
is a uniform time distribution: $F_{\textrm{\tiny Bkg}}(t) = F_{\textrm{\tiny Bkg}}$. 
The background energy distribution $\Lambda_{\textrm{\tiny Bkg}}$ is measured from OFF regions.
$N(\tau_n)$ (resp. $N'$) is the normalization factor of the PDF $P_{\textrm{\tiny Sig}}$ 
(resp. $P_{\textrm{\tiny Bkg}}$) in the range \mbox{$[E_{\rm cut};E_{\rm max}] \times 
[t_{\rm min};t_{\rm max}]$} where the likelihood fit is performed.

Also, the energy resolution $D(E,E_{\rm true})$ is assumed to be perfect in the range 
\mbox{$[E_{\rm cut};E_{\rm max}]$}\footnote{The actual energy resolution is of the order 
of 10~\% in this range.}.
This leads to simplified expressions of $P_{\textrm{\tiny Sig}}(E_i, t_i|\tau_n)$ and 
$P_{\textrm{\tiny Bkg}}(E_i, t_i)$:
\begin{align}
\newline
P_{\textrm{\tiny Sig}}(E_i, t_i|\tau_n) &= \frac{1}{N(\tau_n)} \cdot  A_{\rm eff}(E_i,t_i)
\Lambda_{\textrm{\tiny Sig}}(E_i)F_{\textrm{\tiny Sig}}(t_i - \tau_n \cdot E_i^n)\\
P_{\textrm{\tiny Bkg}}(E_i, t_i) &= \frac{1}{N'} \cdot A_{\rm eff}(E_i,t_i)
\Lambda_{\textrm{\tiny Bkg}}(E_i)F_{\textrm{\tiny Bkg}}
\newline
\end{align}
The best estimate of the dispersion parameter $\widehat{\tau}_n$ is obtained by maximizing 
the likelihood $\mathit{L}(\tau_n)$.

\subsection{Selection cuts}\label{app:selcuts}

Table \ref{tab:cuts} shows the effect of the selection cuts on the number of ON and OFF events. 
Other choices of $E_{\rm min}$ and $E_{\rm cut}$ did not introduce significant changes in the 
final results.
\begin{deluxetable}{lrrr}
\tablecaption{Selections applied to the ON and OFF data sets\label{tab:cuts}}
\tablewidth{0pt}
\tablehead{\colhead{Selection} &\colhead{\# of ON events} &\colhead{Weighted \# of OFF events} 
&\colhead{S/B} }
\startdata
	Total sample                                   &  461 (100 \%) & 144.3 (100 \%) & 2.2 \\
	(1) = Time in 500--8500 s                      & 358 (77.7 \%) & 95.8 (66.4 \%) & 2.7 \\
	(1) and $E$ in 0.3--0.789 TeV                  & 154 (33.4 \%) & 36.3 (25.1 \%) & 3.2 \\
	(1) and $E$ in 0.3--0.4 TeV (Template)         &  82 (17.8 \%) &  14.2 (9.9 \%) & 4.8 \\
	(1) and $E$ in 0.4--0.789 TeV (LH fit)         &  72 (15.6 \%) & 21.9 (15.2 \%) & 2.3 
\enddata
\end{deluxetable}

\subsection{Test of the method, confidence intervals}
\label{annexe:calib}

The method has been tested on Monte Carlo (MC) simulated sets. Each set was composed of 
\mbox{$n_{\rm ON} = 72$} ON events, as in the real data sample: 
\begin{itemize}
\item  $s = 50$ signal events with times following the template light curve (Fig. \ref{fig:template}) 
shifted by a factor $\tau_{n,{\rm inj}} \cdot {\rm E}_i$; energies follow a power law spectrum 
of photon index $\Gamma_{\rm \tiny Sig} = 4.8$, degraded by the acceptance and convolved with
the energy resolution. 
\item $b = 22$ background events with times following a uniform distribution and energies 
drawn from a power law spectrum of index $\Gamma_{\rm \tiny Bkg} = 2.5$, degraded by the 
acceptance and convoluted by the energy resolution. 
\end{itemize}
For a given injected dispersion, the maximum likelihood method is applied to each 
\mbox{MC-simulated} set. 
The initial light curve and energy spectrum were used as templates in the model instead 
of fitting them for each set. 

\begin{figure}[ht]
\center
\includegraphics[width=0.5\textwidth,natwidth=567,natheight=571]{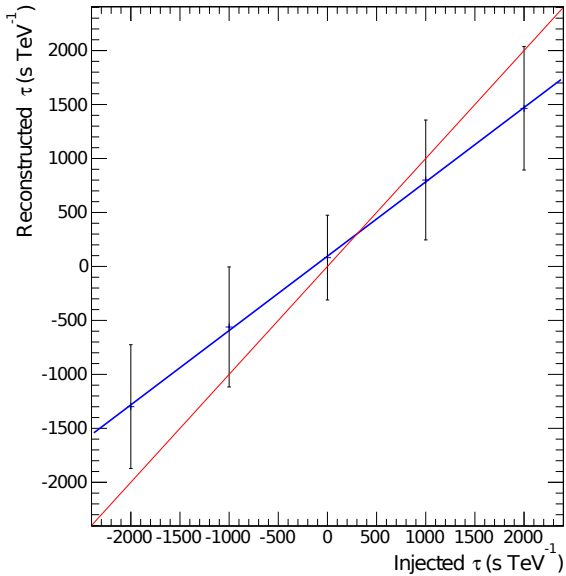}
\caption{Means of the reconstructed dispersion \textit{versus} the real (injected dispersion) 
for the linear case n = 1; for a given injected dispersion, errors bars correspond to the means 
of the distribution of the upper and lower limits (90~\% 2-sided $\simeq$ 95~\% 1-sided). The 
blue line is a linear fit to the points. The red line shows the ideally obtained curve 
$\tau_{\rm recontructed} = \tau_{\rm injected}$ obtained in the case S/B = $\infty$.\label{fig:calib}}
\end{figure}

Figure \ref{fig:calib} shows the means of the reconstructed dispersion \textit{versus} 
the real (injected) dispersion for \mbox{n = 1}; for a given injected dispersion, error 
bars correspond to the RMS of the distribution of the best estimates $\hat{\tau_1}$. The
blue line shows the result of a linear fit. 
The slope roughly corresponds to the percentage of signal in the total ON data set. 
It is due to the loss of sensitivity resulting from the part of the data sets with no dispersion. 
A systematic shift is observed of about 100 s\,TeV$^{-1}$, well bellow 1$\sigma$ value -- the 
RMS of the best estimate distribution is of 361 s\,TeV$^{-1}$. The results in this paper have 
not been corrected for this bias.

\begin{figure}[ht]
\centering
\includegraphics[width=1.02\textwidth]{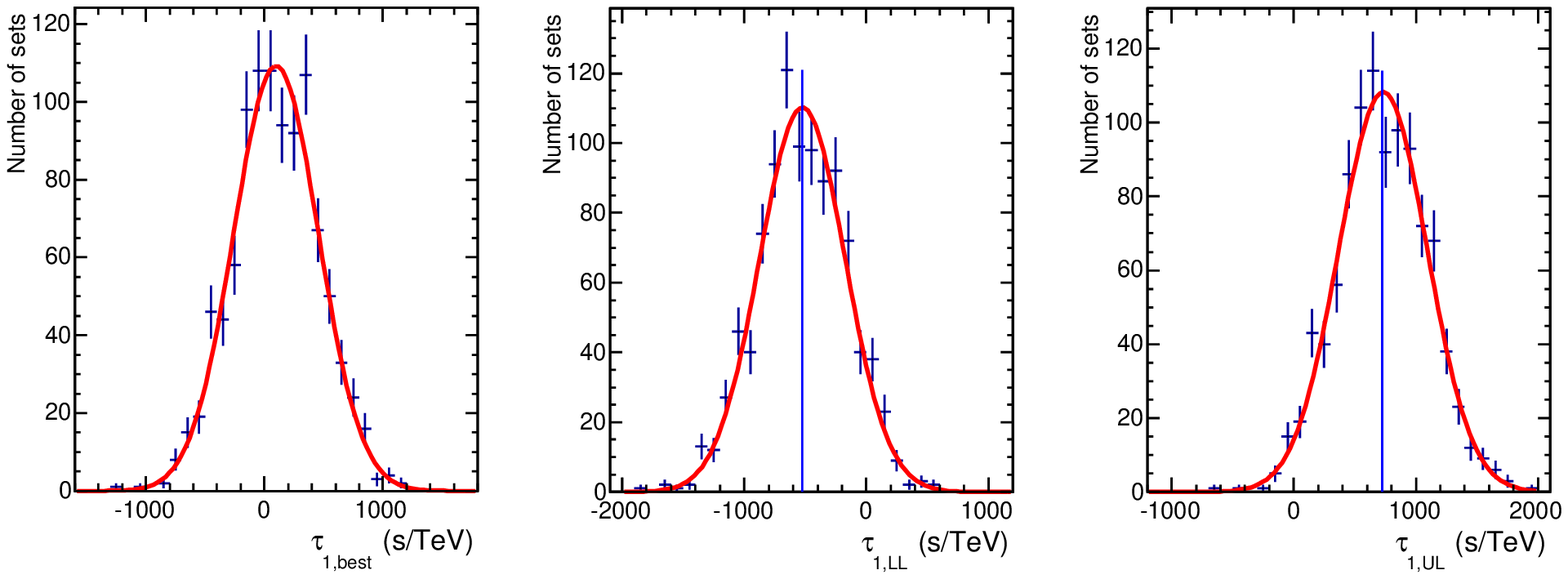}
\includegraphics[width=1.02\textwidth]{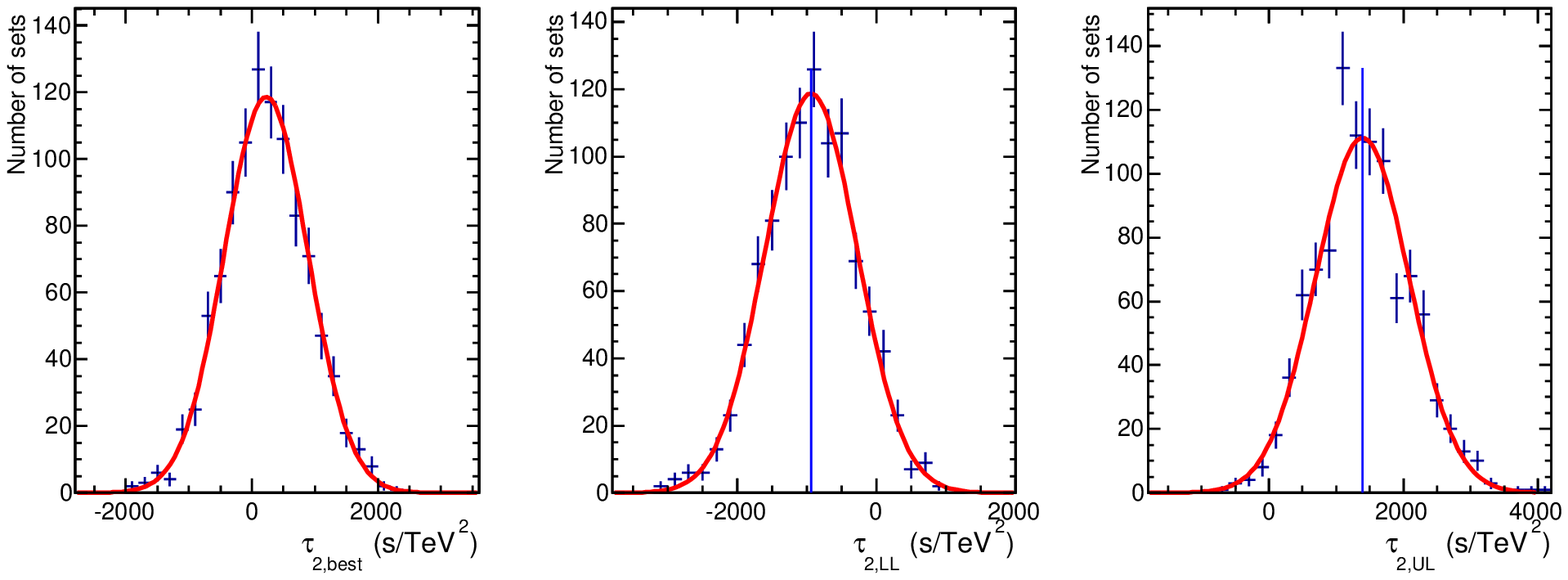}
\caption{Distributions of the best estimates, the 95\% 1-sided lower and upper limits 
from simulations in case of no injected dispersion (\mbox{$\tau_{n, {\rm inj}} = 0$ s\,TeV$^{-n}$}), 
for \mbox{n = 1} (top) and \mbox{n = 2} (bottom); dispersion values are in s\,TeV$^{-n}$. 
The blue vertical line on the LL (resp. UL) distribution shows $LL^{\rm MC}$ (resp. $UL^{\rm MC}$), 
defined as the mean of the distribution.
\label{fig:calib0}}
\end{figure}

The coverage is not necessarily proper, \textit{i.e.} the number of sets for which the 
injected dispersion value $\tau_{\rm inj}$ lies between the set's lower limit (LL) and 
upper limit (UL) does not match the required 95~\% 1-sided confidence level.
The common cut used on the likelihood curves to get the LLs/ULs has been iteratively 
adjusted to ensure a correct statistical coverage: using this new cut, 95~\% of the
realizations provide CIs that include the injected dispersion  $\tau_{n, {\rm inj}}$.
The initial coverage was about 85~\% for a cut on $2\ln \mathit{L}$ of 2.71. The new 
common cut, found iteratively at 3.5, ensures the desired 90~\% 2-sided CL (approx. 95~\% 1-sided CL).
Figure \ref{fig:calib0} shows the distributions of the best estimates, the 95\% 1-sided 
LLs and ULs for \mbox{$\tau_{\rm 1, inj} = 0$ s\,TeV$^{-1}$} (linear case) and 
\mbox{$\tau_{\rm 2, inj} = 0$ s\,TeV$^{-2}$} (quadratic case); the means of the lower 
and upper limit distributions, shown as a blue vertical line, are used to construct 
the ``calibrated confidence interval''.

To get CIs from data, a maximum likelihood method is applied to the original data set 
and gives a best estimate $\tau_{\rm best}^{\rm data}$.
The cut value determined from the simulations to ensure proper coverage is applied on 
the original data set to obtain $LL^{\rm data}$ and $UL^{\rm data}$.
The ``calibrated'' limits $LL^{\rm calib}$ and $UL^{\rm calib}$, combining 
$\tau_{\rm best}^{\rm data}$ from data together with MC results, are taken as
\begin{align}
\newline
LL^{\rm calib} &= \tau_{\rm best}^{\rm data} -|\tau_{\rm best}^{\rm MC} - LL^{\rm MC}| \\ \nonumber
UL^{\rm calib} &= \tau_{\rm best}^{\rm data} +|\tau_{\rm best}^{\rm MC} - UL^{\rm MC}| \nonumber
\newline
\end{align}
with $\tau_{\rm best}^{\rm MC}$, ${\rm LL}^{\rm MC}$ and ${\rm UL}^{\rm MC}$ defined as 
the mean of the per-set best-estimate distribution, LL distribution, and UL distribution 
respectively.

Table \ref{tab:llultau} lists the CIs determined in both ways, i.e., data-only and calibrated 
ones: $LL_n^{\rm data}$ and $LL_n^{\rm calib}$ (resp. $UL_n^{\rm data}$ and $UL_n^{\rm calib}$) 
are compatible within 10~\%.
In this work, calibrated CIs have been used to derive the final lower limits on $\textrm{E}_{\rm QG}$. 
They are preferred over data-only CIs as they provides statistically well defined confidence levels. 
They also ensure coherent comparison with previous published results, e.g. with PKS~2155--304 
by \cite{abramowski_search_2011} and GRB studies by \cite{vasileiou_constraints_2013}.

\begin{deluxetable}{c|ccc|ccc|cc|cc}
\tablecaption{Linear (top) and quadratic (bottom) dispersion parameter; from left to right: 
best estimate, LL and UL from data (cut on likelihood curve), LL and UL from MC simulations 
(means of per-set LL and UL distributions), calibrated LL and UL (combination of data and MC), 
calibrated LL and UL including systematic errors. Dispersion parameters $\tau_{n,best}$, LLs 
and ULs are in s\,TeV$^{-n}$. \label{tab:llultau}}

\tablewidth{0pt}
\tablehead{n & $\tau_{n,best}^{\rm data}$ & $LL_n^{\rm data}$ & $UL_n^{\rm data}$ & 
$\tau_{n,best}^{\rm MC}$ & $LL_n^{\rm MC}$ & $UL_n^{\rm MC}$ & $LL_n^{\rm calib}$ & 
$UL_n^{\rm calib}$ & $LL_n^{\tiny \rm calib}$ & $UL_n^{\rm calib}$
\\ &  &  & & & & & &  &  \multicolumn{2}{c}{with systematics}
}
\startdata
1 & -131.7 & -806.7 & 554.7 & 99.1 & -526.3 & 725.6 & -757.1 & 494.8 & -838.9 & 576.4 \\
2 & -287.5 & -1449.9 & 853.6 & 217.2 & -942.0 & 1395.0 & -1446.7 & 890.3 & -1570.5 & 1012.4 \\ 
\enddata
\end{deluxetable}

\subsection{Estimation of the systematics}
\label{annexe:syst}

Estimations of the systematic effects on the dispersion measurement were performed. It was 
found that the main systematic errors are due to the uncertainties on the light curve 
parametrization. Other sources of systematic errors include the contribution of the 
background, effect of the change of photon index, the energy resolution and the effective 
area determination of the detector.
To study the following four contributions, new simulated data sets have been built, 
each one with different input parameters:
\begin{itemize}
\item background contribution: photons and background events have been reallocated 
within the ON data set in the fit range \mbox{$[E_{\rm cut};E_{\rm max}]$}, introducing 
a $1\sigma$ fluctuation in the number of signal event $s$ in the ON data set;
\item effective area: set to a constant, equal to 120000 m$^2$ for all energies and all 
times, which corresponds to a maximum shift of 10~\% (the actual effective area increases 
with energy);
\item energy resolution: reconstructed energies have been replaced by the true energies; 
this corresponds to a shift of about 10~\% on the reconstructed energy values;
\item photon index: changed by one standard deviation ($\pm 0.25$).
\end{itemize}
For the determination of systematic errors arising from the light curve parametrization, 
the calibration of the confidence intervals has been redone using successively the upper 
$1\sigma$ and the lower $1\sigma$ contours of the template, shown in Fig. \ref{fig:template}.
The change in mean lower and upper limits on the dispersion parameter $\tau_n$ gives an 
estimate of the systematic error associated to each contribution\footnote{In particular the errors 
on the peak positions constitute the most important part of the uncertainty on the template 
light curve contributing to the likelihood fit  -- see previous works, e.g. \citet{abramowski_search_2011}. 
Therefore, the covariance matrix of the fit of the template was studied in detail ; the peak positions 
were varied by values of $\pm 1 \sigma$ extracted from the covariance matrix.
This study led to an increase in overall systematics of the order of 20\% for $\tau_1$ and
40\% for $\tau_2$, and a decrease of maximum 7\% and 2\% of limits on $\textrm{E}_{\rm QG,1}$
and $\textrm{E}_{\rm QG,2}$ respectively.}.
An additional systematic contribution comes from the shift arising from the method found 
with simulation (see Appendix \ref{annexe:calib}).
Table \ref{tab:syst} summarizes all studied systematic contributions. The overall estimated 
systematic error on $\tau_n$ is 330 s\,TeV$^{-1}$ for the linear case (\mbox{n = 1}) and 
555 s\,TeV$^{-2}$ for the quadratic case (\mbox{n = 2}); they were included in the 
calculation of the limits on \eqg\ by adding the statistical and the systematic errors 
in quadrature.
\begin{deluxetable}{ll|ll}
\tablecaption{Summary of all studied systematic contributions. The main systematic errors 
are due to the uncertainties on the light curve parametrization. \label{tab:syst}}
\tablewidth{0pt}
\tablehead{\colhead{}& \colhead{Estimated error} &\colhead{$\tau_1$} &\colhead{$\tau_2$} \\
\colhead{}& \colhead{on input parameters} &\colhead{(s\,TeV$^{-1}$)} &\colhead{(s\,TeV$^{-2}$)}}
\startdata
Background contribution &                      & $< 45$  & $< 80$ \\
Acceptance factors      & 10\%                       &   $< 1$ & $< 1$ \\
Energy resolution       & 10\%                       &   $< 55$& $< 85$\\
Photon index          & 5\%                     &   $< 55$& $< 50$\\
Light curve parametrization &                      &   $< 300$& $< 500$\\
Systematic bias &                      &   $\sim 100$& $\sim 200$\\
\hline
Total: $\sqrt{\sum_i {\rm syst}_i^2}$       &                        & $< 330$  & $< 555$ \\
\enddata
\end{deluxetable}

\bibliography{pg1553flare}
\end{document}